\documentclass[12pt,a4paper]{article}

\usepackage{arxiv}

\usepackage{booktabs}       
\usepackage{graphicx}
\usepackage{subcaption}
\usepackage{xcolor}
\usepackage{comment}
\usepackage{pdflscape}
\usepackage{newtxtext}
\usepackage{newtxmath}
\usepackage{natbib}
\usepackage{hyperref}
\hypersetup{
    colorlinks = true,
    urlcolor   = blue,
    citecolor  = black,
}

\newcommand{\RomanNumeralCaps}[1]

\usepackage{tikz}
\usepackage{tikz-3dplot}
\usetikzlibrary{decorations.pathmorphing,shapes.geometric}

\title{The role of turbulent jet dynamics on contrail formation from conventional and sustainable aviation fuels}

\date{} 					

\author{%
	L. Boscagli$^{1,2}$ \qquad S. Rigopoulos$^{3}$ \qquad
	G. Papadakis$^{1,*}$\\[0.6em]
	\parbox{\dimexpr\textwidth-2\tabcolsep\relax}{\centering\normalfont\small
	$^{1}$ Department of Aeronautics, Imperial College London, London, SW7 2AZ, UK\\
	$^{2}$ Interdisciplinary Center on Sustainability and Climate, Sant’Anna School of Advanced Studies, Pisa, 56127, Italy\\
	$^{3}$ Department of Mechanical Engineering, Imperial College London, London, SW7 2AZ, UK\\[0.3em]
	* Corresponding author: \texttt{g.papadakis@ic.ac.uk}
	}%
}

\renewcommand{\headeright}{}
\renewcommand{\undertitle}{}
\renewcommand{\shorttitle}{Boscagli et al.}

\begin{document}
	\maketitle
	
\begin{abstract}
Large-eddy simulations (LES) coupled with a discretised population balance equation (PBE) are used to quantify the effect of turbulence on the early stages of contrail formation (jet regime) for both conventional (soot-rich) and sustainable aviation fuels (SAFs). SAFs typically have low aromatic content which causes a shift in the dominant nucleation pathway. The contribution from soot (non-volatile particulate matter, \textnormal{nvPM}) is reduced, while the contribution from volatile particulate matter (\textnormal{vPM}) is increased. The computational domain reproduces an experimental jet configuration that replicates the thermodynamic conditions encountered in the upper troposphere. The LES-PBE simulations capture well the jet-core cooling and bracket the experimental values of ice-crystal characteristics (number concentration, mean diameter and mass density) if a range of judicious values for parameters not specified in the experiments are employed. With the aid of one-dimensional, one-way coupled PBE approach, it is shown that soot particle activation can be reliably captured using steady aerosol-jet models. On the other hand, volatile particle activation is significantly affected by large fluctuations in temperature and humidity (and thereby supersaturation). Further downstream, non-linear dynamic coupling between the gas and aerosol phases becomes increasingly important for ice-crystal growth as well. Overall, the results demonstrate that for SAFs, turbulence is an important driver of particle activation and subsequent ice crystal growth and cannot be neglected. This finding motivates more fundamental studies to assess how turbulent jet dynamics modulates contrail formation in realistic engine configurations.
\end{abstract}

\section{Introduction}\label{sec:intro}
Condensation trails (contrails) form downstream of aircraft engines and are a significant, yet still poorly characterised, contributor to radiative forcing \citep{Lee2023}. Under ice-supersaturated conditions, contrails can persist for several hours and evolve into contrail cirrus \citep{Lewellen2014}. Cirrus clouds cause radiative forcing and are one of the most important contributors to aviation-induced climate change. 

A widely used thermodynamic condition to determine the formation of a contrail is the Schmidt-Appleman criterion (SAC), \cite{schumann1996}. Liquid saturation is a necessary and sufficient condition for contrail formation. At a given ambient relative humidity and pressure, SAC provides a threshold atmospheric temperature so that liquid water vapour saturation is reached within the expanding exhaust plume \citep{Karcher1996b}. While SAC provides a thermodynamic threshold, contrail formation is a complex process whose dynamics involves an interaction between microphysical kinetics and fluid dynamics. Understanding this interaction is vital for developing better models and reducing the uncertainty of contrail impact on climate \citep{Lee2023}.

In kerosene-based jet fuels, the typical products of the combustion process include carbon dioxide, water vapour, sulfur and nitrogen oxides, as well as black carbon nanoparticles \citep{Garnier1997}. The latter are also known as soot, and provide the sites for heterogeneous ice nucleation \citep{Murray2022}. One of the most complete descriptions of the nonequilibrium, microphysical pathway is described by \citet{Karcher2015}, also referred to as K15 below. The K15 microphysics model provides a comprehensive pathway for the activation of soot particles (i.e. the point at which they become the core of water droplets), the condensational growth of water droplets, and the formation and depositional growth of ice crystals.  
For kerosene-based (soot-rich) fuels, there is an almost linear dependence among the number of emitted soot particles and the number of ice crystals formed \citep{Karcher2018,Yu2024}.

Sustainable aviation fuels (SAFs) and hydrogen are among the fuel options proposed to mitigate the climate impact of aviation \citep{Jing2022}. Compared to conventional (e.g.\ jet-A) fuels, SAFs have lower aromatic content and thus lower soot particle emissions \citep{Dischl2024}. Depending on the initial aerosol characteristics within the jet plume, one-dimensional contrail models predict that, for SAF, the number concentrations of ice crystals can span up to 3 orders of magnitude range within only a few degrees ($\sim 3\textnormal{ K}$) change in the atmospheric temperature \citep{Ponsonby2025}. This variation stems from a shift in the nucleation pathway. The contribution of soot particles (non-volatile particulate matter - \textnormal{nvPM}) is reduced but that of volatile particulate matter (\textnormal{vPM}) is increased. This is due to the combination of residual sulfuric content in SAF fuel \citep{Karcher2015} and venting of engine lubrication oil droplets \citep{Lee2023} that form new nucleation sites. The activation of vPM in the soot-poor engine exhaust regime is also supported by recent flight tests \citep{Voigt2025_NATURE}, but the uncertainty in the predicted ice crystal numbers remains large. The latter is an important parameter to assess contrail-induced radiative forcing, and the uncertainty can propagate into whole fleet simulations \citep{Grewe2021,Brazzola2025}. From a computational modelling perspective, this indicates the need to further assess the mechanisms driving contrail formation, and in particular aerosol particle activation for SAF and hydrogen.

The first stage of a contrail formation process is referred to as the jet regime \citep{Paoli2016}. In the jet, as well as the vortex regime that follows, the tight coupling between fluid mechanics and contrail microphysics \citep{Scorer1970,Paoli2013} determines the resulting ice particle number density and size distribution, setting the stage for the subsequent evolution in the dissipation and diffusion regimes. Table \ref{tab:contrails_cfd_tools} summarises previous work on contrail modelling. An Eulerian approach is used for the modelling of the gas-phase, and various turbulence modelling strategies have been employed, that include Reynolds (and Favre) averaged Navier Stokes,  RANS \citep{Khou2017,Ramsay2024,Annunziata2025}) and Large Eddy Simulations, LES \citep{Paoli2004,Lewellen2014,Unterstrasser2014,Vancassel2014,Ferreira2026}). RANS modelling approaches typically employ a spatial representation of jet evolution and may also include detailed aero-engine nacelle and airframe geometries. In contrast, owing to the higher computational cost of LES compared to RANS, the former type of studies typically adopts a temporal representation of the jet. In this approach, a slice of the jet flow is discretised and periodic boundary conditions are applied to capture the downstream evolution. This strategy enables the simulation of very long contrails at a relatively low computational cost. For soot-rich plumes with no $\textnormal{vPM}$ contribution, \citet{Afkari2026} found no significant differences in ice crystal characteristics between spatial and temporal representations of the jet. In some cases, precursor simulations are also employed to initialise the jet flow and its subsequent interaction with the aircraft wing-tip vortices \citep{Paoli2013,Lewellen2020,Ferreira2026}.

For particle transport, both Eulerian \citep{Khou2017} and Lagrangian \citep{Paoli2005,Lewellen2014,Unterstrasser2016,Vancassel2014,Ferreira2026,Ramsay2024} approaches have been used. The Lagrangian approach also finds wider use for aerosols in clouds \citep{Bier2024,Yu2024,Shapiro2025,Vals2025}. In principle, due to the low Stokes number, both approaches are valid. The Eulerian approach requires the solution of a population balance equation (PBE), which predicts the evolution of a particle size distribution subject to processes such as particle growth and transport. A general treatment of the PBE and the methods for its solution can be found in \cite{Rigopoulos2024}. The solution approach usually employed for contrails is the method of moments, where only a few moments of the particle size distribution are transported, typically the zeroth and the first \citep{Ferreira2026,Lottermoser2025}. A more detailed approach is to transport a discretised representation of the particle size distribution \citep{Lewellen2020}. The moment approach requires the specification of a size distribution (usually lognormal, \cite{Paoli2016}), while the discretised approach is general, i.e.\ predicts the complete distribution, though this comes at a substantially increased computational cost. 

The fraction of activated aerosols controls the size and concentration of ice crystals at the end of the jet regime \citep{Ponsonby2025}. In the vortex regime, loss of ice crystals due to sublimation and precipitation is typically parameterised as a function of ambient conditions and aircraft and engine configuration \citep{Unterstrasser2020_LCM,Lottermoser2025}. The size and concentration of ice crystals at the end of these regimes are key determinants of the contrail optical thickness, which is required for the evaluation of radiative forcing \citep{Rap2010,SanzMorere2021}. 

\begin{table}
  \begin{center}
\def~{\hphantom{0}}
\begin{tabular}{lllll}
  \multicolumn{1}{c}{\begin{tabular}[c]{@{}c@{}}Reference\\ (tool label)\end{tabular}} &
  \multicolumn{1}{c}{\begin{tabular}[c]{@{}c@{}}Gas-Aerosol\\ framework\end{tabular}} &
  \multicolumn{1}{c}{\begin{tabular}[c]{@{}c@{}}Turbulence\\model\end{tabular}} &
  \multicolumn{1}{c}{\begin{tabular}[c]{@{}c@{}}Ice microphysics\\model\end{tabular}} &
  \multicolumn{1}{c}{\begin{tabular}[c]{@{}c@{}}Contrails\\ regime\\(jet framework)\end{tabular}} \\[20pt] \hline
  \multicolumn{1}{c}{\begin{tabular}[c]{@{}c@{}}\cite{Paoli2004,Paoli2013}\\\cite{Paoli2005}\\(LES-K96)\end{tabular}} &
  \multicolumn{1}{c}{\begin{tabular}[c]{@{}c@{}}Eulerian-\\Lagrangian\end{tabular}} &
  \multicolumn{1}{c}{\begin{tabular}[c]{@{}c@{}}Compressible\\LES\end{tabular}} &
  \multicolumn{1}{c}{\begin{tabular}[c]{@{}c@{}}Based on\\ \citet{Karcher1996b}\end{tabular}} &
  \multicolumn{1}{c}{\begin{tabular}[c]{@{}c@{}}Jet and \\vortex\\(temporal)\end{tabular}} \\[20pt] 
  \multicolumn{1}{c}{\begin{tabular}[c]{@{}c@{}}\cite{Lewellen2020}\\(LES-K15)\end{tabular}} &   
  \multicolumn{1}{c}{\begin{tabular}[c]{@{}c@{}}Eulerian-\\Lagrangian\end{tabular}} &  
  \multicolumn{1}{c}{\begin{tabular}[c]{@{}c@{}}Incompressible,\\variable density\\LES\\(2D RANS for jet)\end{tabular}} &
  \multicolumn{1}{c}{\begin{tabular}[c]{@{}c@{}}Based on\\\cite{Ackerman1995},\\\citet{Karcher2015}\end{tabular}} &
  \multicolumn{1}{c}{\begin{tabular}[c]{@{}c@{}}Jet, vortex, \\dissipation\\and diffusion\\(temporal)\end{tabular}} \\[20pt] 
  \multicolumn{1}{c}{\begin{tabular}[c]{@{}c@{}}\cite{Unterstrasser2014}\\\cite{Unterstrasser2016}\\ \cite{Unterstrasser2020_LCM}\\(EULAG-LCM)\end{tabular}} &     
  \multicolumn{1}{c}{\begin{tabular}[c]{@{}c@{}}Eulerian-\\Lagrangian\end{tabular}} &  
  \multicolumn{1}{c}{\begin{tabular}[c]{@{}c@{}}Anelastic\\LES\end{tabular}} &
  \multicolumn{1}{c}{\begin{tabular}[c]{@{}c@{}}Based on\\\cite{Solch2010}\end{tabular}} &
  \multicolumn{1}{c}{\begin{tabular}[c]{@{}c@{}}Vortex, \\dissipation\\and diffusion\\(n/a)\end{tabular}} \\[20pt] 
  \multicolumn{1}{c}{\begin{tabular}[c]{@{}c@{}}\cite{Vancassel2014}\\(FLUDILES-K96)\end{tabular}} &  
  \multicolumn{1}{c}{\begin{tabular}[c]{@{}c@{}}Eulerian-\\Lagrangian\end{tabular}} &  
  \multicolumn{1}{c}{\begin{tabular}[c]{@{}c@{}}Compressible\\LES\end{tabular}} &
  \multicolumn{1}{c}{\begin{tabular}[c]{@{}c@{}}Based on\\\cite{Karcher1996b}\end{tabular}} &
  \multicolumn{1}{c}{\begin{tabular}[c]{@{}c@{}}Jet and \\vortex\\(temporal)\end{tabular}} \\[20pt] 
  \multicolumn{1}{c}{\begin{tabular}[c]{@{}c@{}}\cite{Khou2017}\\(CEDRE-K96)\end{tabular}} &  
  \multicolumn{1}{c}{\begin{tabular}[c]{@{}c@{}}Eulerian-\\Eulerian\end{tabular}} &  
  \multicolumn{1}{c}{\begin{tabular}[c]{@{}c@{}}Compressible\\RANS\end{tabular}} &
  \multicolumn{1}{c}{\begin{tabular}[c]{@{}c@{}}Based on\\\cite{Karcher1996b}\end{tabular}} &
  \multicolumn{1}{c}{\begin{tabular}[c]{@{}c@{}}Jet and \\vortex\\(spatial)\end{tabular}} \\[20pt]  
  \multicolumn{1}{c}{\begin{tabular}[c]{@{}c@{}}\cite{Ferreira2026}\\(CharLES-K15)\end{tabular}} &  
  \multicolumn{1}{c}{\begin{tabular}[c]{@{}c@{}}Eulerian-\\Lagrangian\end{tabular}} &  
  \multicolumn{1}{c}{\begin{tabular}[c]{@{}c@{}}Compressible\\LES\end{tabular}} &
  \multicolumn{1}{c}{\begin{tabular}[c]{@{}c@{}}Based on\\ \citet{Karcher2015},\\\citet{Bier2022}\end{tabular}} &
  \multicolumn{1}{c}{\begin{tabular}[c]{@{}c@{}} Jet and \\vortex\\(temporal)\end{tabular}} \\[20pt]  
  \multicolumn{1}{c}{\begin{tabular}[c]{@{}c@{}}\cite{Ramsay2024}\\(HYDRA-K96)\end{tabular}} &  
  \multicolumn{1}{c}{\begin{tabular}[c]{@{}c@{}}Eulerian-\\Lagrangian\end{tabular}} & 
  \multicolumn{1}{c}{\begin{tabular}[c]{@{}c@{}}Compressible\\RANS\end{tabular}} &
  \multicolumn{1}{c}{\begin{tabular}[c]{@{}c@{}}Based on\\ \citet{Karcher1996b}\end{tabular}} &
  \multicolumn{1}{c}{\begin{tabular}[c]{@{}c@{}} Jet and \\vortex\\(spatial)\end{tabular}}   
\end{tabular}
\caption{Summary of gas-aerosol coupled computational models for the early stages of contrail formation.}
\label{tab:contrails_cfd_tools}
\end{center}
\end{table}

\textcolor{black}{Table \ref{tab:contrails_fuel_regimes} summarises recent computational studies on contrail formation from alternative aviation fuels (SAF, hydrogen). \citet{Ferreira2025} compares hydrogen with kerosene in terms of the ice emission index; in both cases, ice forms via heterogeneous nucleation on soot-type or ambient particles. \citet{Lottermoser2025} couple LES with a Lagrangian cloud model and detailed ice microphysics to assess how ambient conditions, aircraft and engine configuration, and the apparent ice emission index influence late-stage contrail evolution (diffusion/dispersion regimes); simulations assume ice-supersaturated air and a lognormal initial ice-crystal distribution. \citet{Lewellen2020} couples a discretised population balance with LES to capture the jet and vortex phases for kerosene and SAF fuels.}
 
\begin{table}
  \begin{center}
\def~{\hphantom{0}}
\begin{tabular}{lllll}
\multicolumn{1}{c}{Study} &
  \multicolumn{1}{c}{\begin{tabular}[c]{@{}c@{}}Tool label\end{tabular}} &
  \multicolumn{1}{c}{\begin{tabular}[c]{@{}c@{}}Jet\\framework\end{tabular}} &  
  \multicolumn{1}{c}{\begin{tabular}[c]{@{}c@{}}Aerosols\\type\end{tabular}} &
    \multicolumn{1}{c}{\begin{tabular}[c]{@{}c@{}}Initial aerosols\\distribution\end{tabular}} \\[3pt] \hline
  \multicolumn{1}{c}{\cite{Lewellen2020}} &
  \multicolumn{1}{c}{\begin{tabular}[c]{@{}c@{}}LES-K15\end{tabular}} & 
  \multicolumn{1}{c}{\begin{tabular}[c]{@{}c@{}}Temporal\end{tabular}} &  
  \multicolumn{1}{c}{\begin{tabular}[c]{@{}c@{}}Soot, UFA\\and ambient\end{tabular}} &
    \multicolumn{1}{c}{\begin{tabular}[c]{@{}c@{}}Monodisperse\\and bimodal\end{tabular}} \\  
  \multicolumn{1}{c}{\cite{Ferreira2025}} &
  \multicolumn{1}{c}{\begin{tabular}[c]{@{}c@{}}CharLES-K15\end{tabular}} &
  \multicolumn{1}{c}{\begin{tabular}[c]{@{}c@{}}Temporal\end{tabular}} &  
  \multicolumn{1}{c}{\begin{tabular}[c]{@{}c@{}}Soot\\and ambient\end{tabular}} &
    \multicolumn{1}{c}{\begin{tabular}[c]{@{}c@{}}Monodisperse\end{tabular}} \\ 
  \multicolumn{1}{c}{\cite{Lottermoser2025}} &
  \multicolumn{1}{c}{\begin{tabular}[c]{@{}c@{}}EULAG-LCM\end{tabular}} & 
  \multicolumn{1}{c}{\begin{tabular}[c]{@{}c@{}}Temporal\end{tabular}} &  
  \multicolumn{1}{c}{\begin{tabular}[c]{@{}c@{}}Ice\\crystals\end{tabular}} &
    \multicolumn{1}{c}{\begin{tabular}[c]{@{}c@{}}Lognormal\end{tabular}} 
\end{tabular}
\caption{Summary of computational studies with (resolved) gas-phase turbulence and aerosol kinetics for contrails formation from alternative fuels.}
\label{tab:contrails_fuel_regimes}
\end{center}
\end{table}  

Despite the growing number of scale-resolving simulations, the effect of jet turbulence on the activation fraction of ice nucleating particles (INPs), especially in the presence of vPMs, remains largely unexplored. The sensitivity of aerosol activation to temperature and humidity means that turbulent fluctuations may have an important impact on activation and the subsequent development of the ice phase. Large fluctuations in temperature and humidity (and thereby supersaturation) are expected to occur during the spatial evolution of the jet due to the high shear between the jet and the surrounding atmosphere.  Prior work has quantified the sensitivity of the number of formed ice crystals to changes in equilibrium ambient temperature \citep{Ponsonby2025}. A larger fraction of active INPs increases the number of ice crystals and reduces particle radii, therefore affecting contrail optical thickness. 

Within the above context, the central objective of the  paper is to improve the understanding of the interaction between turbulence and contrail formation processes. This is achieved by coupling gas-phase LES with discretised PBE that includes detailed microphysics (K15 activation pathway) as well as vPM; for the latter the parcel model of \cite{Ponsonby2025} is adopted. The coupled approach is validated against the lab experiments of \cite{Fagan2024} that can replicate the thermodynamic conditions of the upper troposphere. The inclusion of vPM is crucial for SAFs, as \cite{Ponsonby2025} have demonstrated that even small variations in ambient temperature, on the order of a few degrees, can alter ice‑crystal number concentrations for SAF by orders of magnitude. The modelling approach simulates the complete jet evolution, including the jet core and its breakup, and the region of intense turbulent kinetic energy production. This allows a detailed investigation of how these processes shape the early‑stage contrail development.  The resulting analysis quantifies the role of turbulence on particle activation and provides guidance for future studies aimed at assessing how turbulent jet dynamics may influence radiative‑forcing estimates from contrails produced by SAF‑fuelled aircraft.

The rest of the manuscript is organised as follows. Sections \ref{sec:methods} and \ref{sec:1d_pbe_method} introduce the  computational frameworks, 3D and 1D respectively. Section \ref{sec:configuration} details the flow configuration, while section \ref{sec:results} presents comparison with experimental data (sections \ref{sec:results_les} and \ref{sec:results_les_pbe}), and a detailed analysis of the interaction between fluid mechanics and aerosol microphysics (sections \ref{sec:results_spatial_temporal_effect}, \ref{sec:fluid_dynamics_activation} and \ref{sec:fluid_dynamics_crystal_growth}). Finally, section \ref{sec:conclusions} summarises the key conclusions and impact of this work. 

\section{Computational framework}\label{sec:methods}

\subsection{Governing equations and numerical solution method}
The density-weighted, spatially-filtered compressible Navier-Stokes equations are solved. They comprise the conservation equations of mass \eqref{eq:continuity} and momentum \eqref{eq:momentum}, as well as the transport equations for gas chemical species $\alpha$, \eqref{eq:species} and enthalpy \eqref{eq:enthalpy},

\begin{equation}
\frac{\partial\hat{\rho}}{\partial t} + \frac{\partial\hat{\rho} \tilde{u_i}}{\partial x_i} = \dot{\hat{\omega}} \, ,
\label{eq:continuity}    
\end{equation}

\begin{equation}
\frac{\partial\hat{\rho} \tilde{u_i}}{\partial t} + 
\frac{\partial\hat{\rho} \tilde{u_i} \tilde{u_j}}{\partial x_j} = -\frac{\partial\hat{p}}{\partial x_i} +
\frac{\partial}{\partial x_j} \left( \hat{\mu} \left(\frac{\partial \tilde{u}_i}{\partial x_j} + \frac{\partial \tilde{u}_j}{\partial x_i} \right)\right) - \frac{\partial \tau_{ij}}{\partial x_j} \, ,
\label{eq:momentum}    
\end{equation}

\begin{equation}
\frac{\partial\hat{\rho} \tilde{Y}_{\alpha}}{\partial t} + 
\frac{\partial\hat{\rho} \tilde{u_i} \tilde{Y}_{\alpha}}{\partial x_j} = 
\frac{\partial}{\partial x_j} \left( \frac{\hat{\mu}}{\sigma}\frac{\partial \tilde{Y}_{\alpha}}{\partial x_i}\right) - \frac{\partial J_{\alpha,i}}{\partial x_i} + \dot{\hat{\omega}}_{\alpha}  \, ,
\label{eq:species}    
\end{equation}

\begin{equation}
\frac{\partial\hat{\rho} \tilde{h}}{\partial t} + 
\frac{\partial\hat{\rho} \tilde{u_i} \tilde{h}}{\partial x_j} = 
\frac{\partial}{\partial x_j} \left( \frac{\hat{\mu}}{\sigma}\frac{\partial \tilde{h}}{\partial x_i}\right) - \frac{\partial J_{h,i}}{\partial x_i} + \dot{\hat{q}} .
\label{eq:enthalpy}    
\end{equation}

In the above set,  $\hat{(\cdot)}$ denotes a spatially-filtered variable, $\tilde{(\cdot)}$ a density-weighted filtered variable (defined as $\tilde{f}(x,y,z;t)=\widehat{\rho f}/\hat{\rho}$), $Y_{\alpha}$ the mass fraction of species $\alpha$, $\tau_{ij} \left( =\hat{\rho}\left(\widetilde{u_i u_j} -\tilde{u}_i \tilde{u}_j \right) \right)$ the subgrid scale stress tensor, $\mu$ the molecular viscosity. In the rest of the manuscript, the symbols $\hat{(\cdot)}$ and $\tilde{(\cdot)}$ are dropped for a more concise notation. In \eqref{eq:momentum}, $\tau_{ij}$ is determined with the dynamic \cite{Lilly1992} model. In \eqref{eq:species} and \eqref{eq:enthalpy}, Fick's and Fourier's laws are used to approximate the diffusive fluxes of the species and heat respectively. The Schmidt number is assumed constant for all species and equal to the Prandtl number ($\sigma=0.7$),  \cite{Jones2012}. $J_{\alpha,i}$ and $J_{h,i}$ are the subgrid scale fluxes for the species mass fraction and enthalpy respectively; they are computed as detailed in \citet{Jones2007}. Finally, $\dot{\omega}$ and $\dot{q}$ denote the volumetric mass-transfer rate and the net rate of heat addition, respectively, as determined by the ice microphysics (see Section \ref{sec:microphysics}). They represent the water vapour sink and the heat released into the gas phase due to condensation/deposition, respectively. The precise definition of $\dot{\omega}$ is provided later in this section.

For the modelling of the transport and kinetics of the ice nucleating particles (INPs), a discretised Population Balance Equation (PBE) approach is employed, in a formulation akin to \cite{Tang2020,Tang2022,Tang2023}; see also \cite{Rigopoulos2024}, section 4.3. The PBE is volume-based, and  in accordance with previous computational studies of the jet and vortex regimes  \citep{Paoli2004,Ferreira2026}, a non-inertial particle approximation is used, where particles are assumed to follow the fluid. This modelling assumption was verified a posteriori by evaluating the Stokes number ($St$). The latter was computed using the Stokes approximation for the particle relaxation time and both the Kolmogorov and eddy turnover timescales for the flow (accounting for both resolved and sub-grid scale contributions). The values were $St\sim\mathcal{O}(10^{-2}-10^{-6})$ within the jet plume, small enough to justify the aforementioned assumption. 

The PBE equations take the form, 
\begin{equation}
\frac{\partial n_i}{\partial t}
+
\frac{\partial \left(u_{j} n_i\right)}{\partial x_j}
=
\frac{\partial}{\partial x_j}
\left(
\frac{\mu}{\rho\,\sigma}
\frac{\partial n_i}{\partial x_j}
\right)
-
\frac{\partial \left(G(v)n_i\right)}{\partial v},
\label{eq:pbe_step1}
\end{equation}
where $v$ is the particle volume and $n_{i}$ is the number concentration per unit of bin width ($dv$). The volume growth function $G(v)$ is defined as
\begin{equation}
G(v) = 3 \left(\frac{4\pi}{3}\right)^{1/3} v^{2/3} \frac{dr}{dt}
\label{eq:growth_rate}    
\end{equation}
where $dr/dt$ is the radial growth rate and it is computed with the microphysical model detailed in section \ref{sec:microphysics}. 

Similar to previous work \citep{Paoli2004,Paoli2013}, mass coupling between the gas and particulate phases is achieved through the source term $(\dot{\omega})$ in the continuity \eqref{eq:continuity} and transport equation \eqref{eq:species} for water vapour concentration  ($Y_{\alpha}=Y_{H_2O}$); this term accounts for supersaturation consumption due to the growth of particles. More specifically, $\dot{\omega}$ is determined by the rate of change of the first moment $M_1$ \citep{Rigopoulos2024},
\begin{equation}
    \dot{\omega} = -\frac{dM_1}{dt} = -\frac{d}{dt}\int v \, n(v) \,dv = \int v \left( \frac{\partial G(v)n(v)}{\partial v}\right)dv \, .
\end{equation}
In addition, a source term ($\dot{q}$) is added to the transport equation for enthalpy \eqref{eq:enthalpy} to account for latent heat release. This term is positive when vapour condensation and ice deposition release heat into the gas phase and is defined as $\dot{q}=-\dot{\omega}l_{c/d}$, where $l_{c/d}$ denotes the latent heat associated with condensation (subscript c) and deposition (subscript d), and is computed as detailed in Appendix \ref{app:thermo_correlations}. The latent heat release associated with freezing is neglected, as freezing is assumed to occur instantaneously \citep{Karcher2015}.

The parallel, multi-block LES solver BOundary Fitted Flow INtegrator (BOFFIN, \cite{Jones2012}) is used to solve the set of equations \eqref{eq:continuity}-\eqref{eq:enthalpy}. BOFFIN employs a nominally second-order finite volume discretisation method. It is based on a low Mach number, pressure-based formulation, with collocated variable  arrangement. Second-order central differences are used for the discretisation of the convective and diffusive terms of the momentum equation, and for the diffusive terms of the scalars. The convective terms of the scalars are discretised with a Total Variation Diminishing (TVD) scheme \citep{VanLeer1974}. The implicit Crank-Nicholson method is used for time integration.

Equation \eqref{eq:pbe_step1} is advanced in time using a fractional step method (see \cite{Rigopoulos2024}, section 4.4.4). This method splits the solution in two steps. In the first step, a partial differential equation (PDE) for a scalar is solved, and in the second a spatially homogeneous hyperbolic PDE that contains only the growth term, 
\begin{equation}
    \frac{\partial n_{i}}{\partial t} + \frac{\partial  G(v) n_{i}}{\partial v}=0,
    \label{eq:pbe_step2}
\end{equation}
is solved. This approach facilitates the integration of the PBE solver CPMOD \citep{Rigopoulos2024} into the the Navier-Stokes solver BOFFIN. The hyperbolic PBE is discretised with a finite volume approach, and $G(v)$ is computed at each interface to conserve the zeroth moment of the volume distribution. A TVD scheme \citep{Qamar2006} is employed to compute the fluxes, while an explicit forward Euler integration scheme is used for temporal integration. 

\subsection{Ice microphysics model} \label{sec:microphysics}
The ice-microphysics model, denoted as K15, is based on \cite{Karcher2015}. Its implementation closely follows the work of \cite{Bier2022} and \cite{Ponsonby2025}. The activation of the aerosol particles is determined by the vapour saturation ratio relative to liquid ($S_{v}$), which is defined as
\begin{equation}
    S_v = \frac{p_v}{p_v^{\mathrm{liq}}} = \frac{pX_{H_2O}}{p_v^{\mathrm{liq}}},
    \label{eq:Sv}
\end{equation}
where $p$ is the gas static pressure, $p_v$ and $X_{H_2O}$ are the water vapour partial pressure and molar fraction respectively, and $p_v^{liq}(T)$ is the saturation partial pressure relative to liquid;  the mathematical expression for $p_v^{liq}(T)$ can be found in  appendix \ref{app:thermo_correlations}. Aerosol particles activate when $S_v$ overcomes the equilibrium saturation ratio $S_K$, which is computed based on the $\kappa-$K\"{o}hler equation \citep{Petters2007}, 
\begin{equation}
    S_K = \frac{r_w^3-r_d^3}{r_w^3-r_d^3(1-\kappa)}\exp{\left(\frac{2\sigma_w M_w}{r_w\mathcal{R} T \rho_w}\right)},
    \label{eq:kappa_kohler}
\end{equation}
where $r_w$ and $r_d$ are the (wet) particle and (dry) core radii respectively, $\sigma_w$ is the surface tension of a water droplet \citep{Pruppacher1997},  $\kappa$ is the hygroscopicity of the core, $T$ is the surrounding gas temperature, $\rho_w$ and $M_w$ are the density and mass molar weight of water respectively, and $\mathcal{R}$ is the universal gas constant ($\approx8314.14$ J/kg/kmol). 

Equation \eqref{eq:kappa_kohler} combines mathematically the Kelvin (curvature) and Raoult (solute) mechanisms that compete to determine the equilibrium vapour pressure over a solution droplet for a prescribed local temperature \citep{Lohmann2016}. The Kelvin term accounts for the curvature of the droplet surface. Since creating additional surface area requires energy, small droplets (which have a larger surface-to-volume ratio)  exhibit a higher equilibrium vapour pressure than larger droplets or a flat water surface. On the other hand, the Raoult term accounts for the presence of the dissolved solute, parameterised by the hygroscopicity parameter $\kappa$ \citep{Petters2007}. Solute molecules reduce the mole fraction of water at the droplet surface, thereby lowering the equilibrium vapour pressure relative to pure water and promoting condensation. The $\kappa-$K\"{o}hler theory therefore describes droplet activation as the balance between the curvature effect (which inhibits growth by increasing the equilibrium saturation ratio) and the solute effect (which favours growth by reducing it).
\begin{figure}
  \centering 
  \includegraphics[trim={0cm 0cm 0cm 0cm},clip,width=0.5\textwidth]{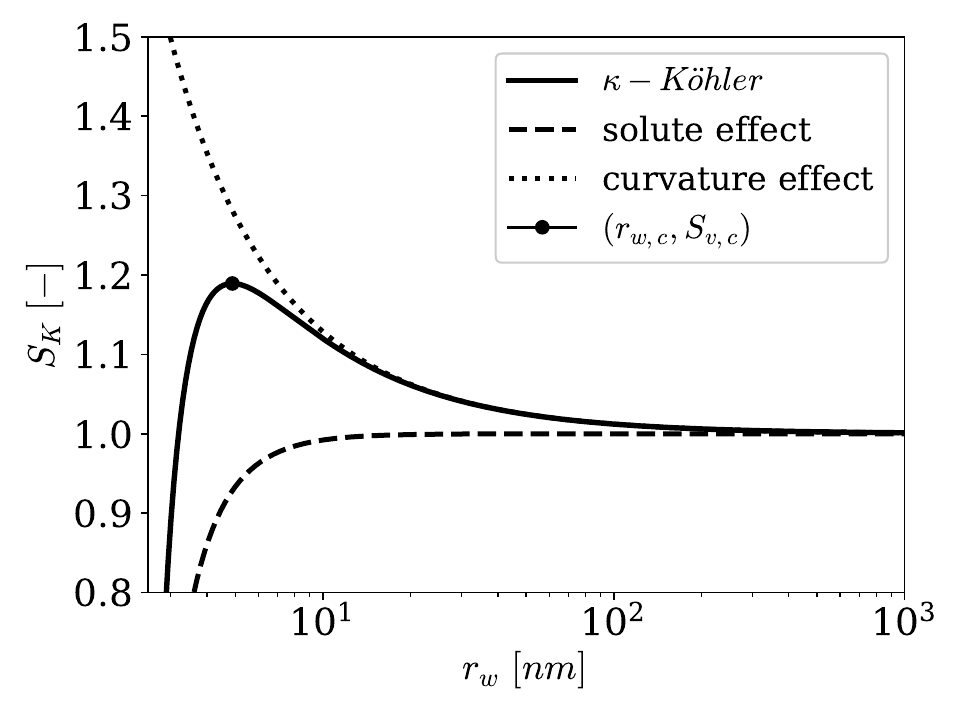}
  \caption{Schematic diagram of the $\kappa-$K\"{o}hler theory, depicting the solute and curvature effects on the equilibrium saturation ratio, $S_K$. Plot for $(r_d,\kappa)=(2.5\,nm, 0.5)$; $T=273.15\textnormal{ K}$.}
  \label{fig:kappa_kohler}
\end{figure}

Since a volume-based PBE is used in this work, the radii are determined from the particle volume assuming spherical shape. The critical particle radius and equilibrium saturation ratio $(r_{w,c},S_{v,c})$, marked with a dot in figure \ref{fig:kappa_kohler}, are determined by solving $dS_K/dr_w=0$, which reduces to finding the roots of
\begin{equation}
f(r_{w,c}) =
-\frac{2 \sigma M_w}{R T \rho_w} r_{w,c}^6
+ 3 \kappa r_d^3 r_{w,c}^4
+ r_d^3 \frac{2 \sigma M_w}{R T \rho_w} (2-\kappa)
- (1-\kappa)\frac{2 \sigma M_w}{R T \rho_w} r_d^6
= 0  \, .
\end{equation}
This equation is solved numerically via the bisection method for each dry nuclei radius ($r_d$) in every cell of the computational domain and at every timestep. $S_{v,c}$ is computed by inserting $r_{w,c}$ in \eqref{eq:kappa_kohler}. If $S_v>S_{v,c}$, then all the particles with a radius greater than $r_{w,c}$ activate into water droplets.

Once activated, the condensational growth of the particles is driven by the difference between the local liquid water vapour saturation and the equilibrium saturation ratio $S_v - S_K$, see \cite{Bier2022}. The mass growth rate of a spherical water droplet with radius $r$ is given by
\begin{equation}
    \frac{dm_w}{dt} = 4\pi r
\frac{p_{v} - S_K p_{v}^{liq}}{F_M/\beta_m + F_H/\beta_t},
\label{eq:water_dmdt}
\end{equation}
where $F_M$ and $F_H$ are the mass and heat diffusion terms respectively, $\beta_m$ and $\beta_t$ are the correction factors due to the large Knudsen number \citep{Bier2022},  and $p_{v}=p X_{H_2O}$ is the water vapour partial pressure, see \eqref{eq:Sv}. The radial growth rate is obtained from the mass growth rate as:
\begin{equation}
    \frac{dr}{dt} = \frac{1}{4\pi r^2 \rho_{w}} \left(\frac{dm_w}{dt}\right),
    \label{eq:water_drdt}
\end{equation}
where $\rho_{w}=997.0 \textnormal{ kg/m}^3$ is the constant density of a supercooled water droplet. This expression is inserted in \eqref{eq:growth_rate} to compute the volume growth function $G(v)$. Upon activation into water droplets, the dry core of the particle is neglected. This is justified by the considerably smaller dry core volume compared to the wet volume, and it enables the modelling of activation of both vPM and nvPM using a single PBE. 

Based on \cite{Karcher2015}, (instantaneous) homogeneous freezing theory is used to determine the freezing temperature $T_{frz}$ of the super-cooled water droplets that formed upon the activation and condensational growth of INPs. $T_{frz}$ is a function of the jet cooling rate and the liquid water volume surrounding the dry core (further details can be found in appendix \ref{app:ice_microphysics}). When the ambient, local temperature is below  $T_{frz}$, the supercooled water droplet instantaneously freezes and continues to grow with depositional growth. This growth is driven by the difference between $S_v$ and the saturation ratio relative to ice $S_{v,i}$ ($=p_v/p_v^{\mathrm{ice}}$). The mass growth rate of a spherical ice crystal is computed as:
\begin{equation}
    \frac{dm_{ice}}{dt} = 4\pi \, r \,
D_{v} \, \beta_v^{-1}\frac{p_{v} - S_K p_{v}^{ice}}{\left( \frac{D_{v} \, \beta_v^{-1} \, l_d \, S_v \, p_{v}^{ice}}{k_{air} \, \beta_k^{-1} \, \beta_v^{-1} \,T}\right)
\left(\frac{l_d}{R_v T}-1\right) + R_v T}
\label{eq:ice_dmdt}
\end{equation}
where $D_{v}(T,p)$ is the diffusion coefficient between air and water vapour, and  $\beta_k$ and $\beta_v$ are correction factors for heat and mass diffusion respectively (computed as in \citet{Bier2022}), which are needed due to the large Knudsen number of the aerosol particles; $R_v$ is the specific gas constant for water vapour, $k_{air}(T)$ is the thermal conductivity of air, and $l_d$ is the latent heat of deposition. The radial growth rate is obtained from the mass growth rate as follows:
\begin{equation}
    \frac{dr_{ice}}{dt} = \frac{1}{4\pi r^2 \rho_{ice}} \left(\frac{dm_{ice}}{dt}\right) \, ,
\label{eq:ice_drdt}
\end{equation}
where $\rho_{ice}=917.0 \textnormal{ kg/m}^3$ is the constant density of the ice crystal. Again, this is inserted in \eqref{eq:growth_rate} to compute the volume growth function $G(v)$.

Liquid and ice saturated conditions are computed based on the experimental correlations for liquid ($p_{v}^{\mathrm{liq}}$) and ice ($p_{v}^{\mathrm{ice}}$) saturation, respectively, and the latent heat of condensation and deposition ($l_{c/d}$) is computed from correlations that use the local temperature. Further details can be found in appendix \ref{app:thermo_correlations}.

\subsection{Ice crystal characteristics} \label{sec:methods_data_analysis}
The ice crystal characteristics are determined from the (conditional) zero-th ($M_{0,ice}$) and first ($M_{1,ice}$) moments of the instantaneous aerosol particles distribution, \begin{equation}
    M_{0,ice}(\textbf{x};t) = \int_{v_0}^{v_{N_{bins}-1}} \delta_{i,ice} n_i \, dv \, 
    \label{eq:M0_ice}
\end{equation}
and
\begin{equation}
    M_{1,ice}(\textbf{x};t) = \int_{v_0}^{v_{N_{bins}-1}} \delta_{i,ice} v_i n_i \, dv \, .
    \label{eq:M1_ice}
\end{equation}
The integration bounds $v_0$ and $v_{N_{bins}-1}$ are the volume size of the particles in the first and last bin of the distribution, respectively; $\delta_{i,ice}$ is the Kronecker delta (1 if the particles in the bin are ice crystals and 0 if they are either water droplets or inactive INPs). The status of the aerosol particles is determined from the ice-microphysics model outlined in section \ref{sec:microphysics}. The instantaneous total number concentration of ice particles $N_{ice}(\textbf{x};t)$ is equal to $M_{0,ice}$ and the mean diameter $d_{mean,ice}$ is determined from the mean volume (assuming spherical particles) as
\begin{equation}
    d_{mean,ice}(\textbf{x};t) = \left(\frac{6}{\pi}\frac{M_{1,ice}}{M_{0,ice}}\right)^{1/3} \, .
    \label{eq:d_ice_mean}
\end{equation}

\subsection{Metrics to assess the effect of turbulence}
To assess the effect of turbulence on the activation of INPs, activation diagnostics are computed by comparing the local supersaturation $s(\textbf{x};t)=S_v(\textbf{x};t)-1$ with the $\kappa-$K\"{o}hler critical equilibrium saturation ratio,  $S_{v,c,i}(T)$. For each nuclei $i$ with dry radius $r_{d,i}$ and hygroscopicity $\kappa_i$, the instantaneous, local activation indicator, $I_i(\textbf{x};t)$, is defined as
\begin{equation}
    I_i (\textbf{x};t)=
    \begin{cases}
        1, & S_v \ge S_{\mathrm{v,c},i}(T),\\
        0, & \text{otherwise}.
    \end{cases}
\end{equation}
Following activation, which is considered irreversible, the indicator remains equal to one. The nuclei are grouped as vPM ($\kappa_i=0.5$) and nvPM ($\kappa_i=0.005$). For each particle class $c\in\{\mathrm{vPM},\mathrm{nvPM}\}$ point metrics are computed as
\begin{equation}
f_c(\mathbf{x},t)=\frac{1}{N_c}\sum_{i\in c} I_i(\mathbf{x},t),    
\end{equation}
where $N_c$ is the number of nuclei in class $c$. 

A conditional joint probability density function (PDF) between $T$ and $S_v$, $p(T,S_v\mid z,\ S_v>1)$, is computed via a 2D histogram. To improve statistical convergence, at each streamwise location $z$, the samples are enriched by exploiting the rotational symmetry of the flow (see details in section  \ref{sec:configuration}). 

The instantaneous and time-average activation probabilities at streamwise location $z$, $P_c^{\mathrm{inst}}(z)$ and $P_c^{\mathrm{mean}}(z)$ respectively, are defined as,
\begin{equation}
P_c^{\mathrm{inst}}(z)=\left\langle f_c^{\mathrm{inst}}\right\rangle_c,\qquad
P_c^{\mathrm{mean}}(z)=\left\langle f_c^{\mathrm{mean}}\right\rangle_c.
\end{equation}
The operator $\langle\cdot\rangle_c$ evaluates the arithmetic mean over $(r,\theta;t\mid z )$ samples,
\begin{equation}
\left\langle g \right\rangle_c
=
\frac{1}{\mathcal{N}_s(z)}\sum_{i\in\mathcal{S}(z)} g_i, 
\end{equation}
where $\mathcal{S}(z)$ is the set of pooled spatio-temporal samples at the streamwise station $z$, and $\mathcal{N}_s(z)$ is the number of samples in that set. 

The contribution of turbulent fluctuations is quantified from
\begin{equation}
  \Delta P_c(z)=P_c^{\mathrm{inst}}(z)-P_c^{\mathrm{mean}}(z),  
\end{equation}
with $\Delta P_c>0$ ($<0$) indicating fluctuation-enhanced (suppressed) activation relative to the time-average. To quantify the effect of the dry radius of the INPs, binwise metrics are also computed as
\begin{equation}
   P_i(z)=\langle I_i\rangle_i. 
\end{equation}
where $\langle\cdot\rangle_i$ denotes averaging over pooled samples of dry radius $r_{d,i}$ at fixed $z$.

\section{One-dimensional PBE model} \label{sec:1d_pbe_method}

To aid the analysis and interpretation of the LES-PBE results, equation \eqref{eq:pbe_step2} is also solved along multiple streamlines using a prescribed time-dependent temperature and dilution model. This simplified formulation is referred to as the 1D-PBE model, because the equation is solved only in the particle-size dimension $v$. The jet temperature and mixing timescale are modelled based on \cite{Karcher2015} with initial (jet, $j$) and ambient ($a$) conditions determined from the LES-PBE data. To obtain the mixing line from the scale-resolving simulations, the local molar fraction of $\mathrm{CO_2}$ is used as a passive scalar tracer to quantify dilution. For each streamline, the local jet fluid fraction is computed as
\begin{equation}
    f_{\mathrm{j}}(z) = \frac{X_{\mathrm{CO_2}}(z)}{X_{\mathrm{CO_2},j}},
\end{equation}
where $X_{\mathrm{CO_2}}(z)$ is the local $\mathrm{CO_2}$ molar fraction (the dependence on x and y is dropped for a more concise notation), $X_{\mathrm{CO_2},j}$ is the value at $z=0$ on the jet centreline.  Using $f_{\mathrm{j}}(z)$, the partial pressure of water vapour and temperature (that define the mixing line) can be obtained, 
\begin{align}
    p_{v}^{\mathrm{mix}}(z) &= f_{\mathrm{j}}(z)\, p_{v,\mathrm{j}} + [1 - f_{\mathrm{j}}(z)]\, p_{v,\mathrm{a}}, \\
    T^{\mathrm{mix}}(z)   &= f_{\mathrm{j}}(z)\, T_{\mathrm{j}} + [1 - f_{\mathrm{j}}(z)]\, T_{\mathrm{a}},
\end{align}
where the subscripts $j$ and $a$ denote jet exit and ambient values, respectively. For each streamline where SAC is satisfied, the slope of the mixing line $G^{\mathrm{mix}}$ (with units [$\mathrm{Pa \,K^{-1}}$]) is then obtained by least-square fit to the linear relation 
\begin{equation}
    p_{v}^{\mathrm{mix}}(z) - p_{v,\mathrm{a}} = G^{\mathrm{mix}}\, [T^{\mathrm{mix}}(z) - T_{\mathrm{a}}].
\end{equation}
Generalised approaches to parametrise $p_v^{\mathrm{mix}}$ that adapt to novel engine and aircraft configurations exist \citep{Richardson2025}, but the approach used in this work is consistent with standard methods in contrail and jet plume studies \citep{schumann1996,Karcher2015,Ponsonby2025}. The values of $G^{\mathrm{mix}}$ derived from the LES-PBE data are used for parametric investigations using the 1D-PBE model. The investigations are used to examine how the thermodynamic history of the jet affects particle activation and aid the qualitative analysis of the LES-PBE results. In the rest of the manuscript, the 1D-PBE results are clearly marked with the superscript $(\cdot)^{1D-PBE}$.

\section{Flow configuration, computational domain, and grid  parameters}\label{sec:configuration}
The NASA Particulate Aerosol Laboratory (PAL) is a facility that was set up to study the early stages of contrail formation in a controlled environment \citep{Wong2013}. The facility can replicate the thermodynamic conditions encountered by an aircraft flying in the upper troposphere. Recent experimental investigations have assessed contrail formation for various fuels, including soot-rich fuels such as n-butanol and Jet-A, and sustainable aviation fuels, such as HEFA-SPK  \citep{Fagan2024}. This configuration therefore provides a useful test case for validation of the computational method presented in section \ref{sec:methods}. The validated methodology will then be used to assess the effect of turbulence on the activation of volatile and non-volatile particles and the ice crystal characteristics.  

\begin{figure}
\centering
\includegraphics[width=0.75\textwidth]{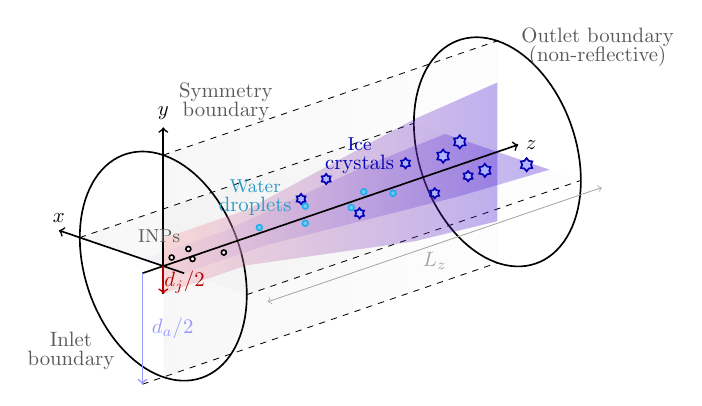}
\caption{Schematic diagram of the computational domain.}
\label{fig:les_pbe_domain}
\end{figure}

A schematic diagram of the computational domain that shows basic dimensions, boundary conditions, and the changing status of the aerosol particles is depicted in figure \ref{fig:les_pbe_domain}. Cylindrical coordinates $(r, \theta, z)$ and Cartesian coordinates $\textbf{x}=(x,y,z)=(r\cos\theta, r\sin\theta, z)$ are used interchangeably. The velocity components along the $(x,y,z)$ directions are denoted as $(u,v,w)$ respectively. The jet diameter $d_j=0.018\textnormal{ m}$ and the domain size $d_a=0.6\textnormal{ m}, \, L_z=1.4\textnormal{ m}$ are based on the experimental configuration of \cite{Fagan2024}. The computational grid consists of approximately 9.1 million cells. The total number of grid nodes along the radial, azimuthal and streamwise directions are $(N_{grid,r},N_{grid,\theta},N_{grid,z})=(77,240,700)$. 

In agreement with previous studies \citep{Paoli2004}, the jet  exit velocity and temperature profiles are blended to the free-stream using a hyperbolic-tangent function. A random, 1\% streamwise velocity perturbation is imposed at $w(r)$ at the free-stream. The in-plane ($x,y$) velocity components ($u,v$) and turbulence intensity are set to zero. At the outlet, non-reflective characteristic boundary conditions \citep{Lodato2008} are imposed, while symmetry conditions are applied on the outer cylindrical boundary (equivalent to a slip, adiabatic wall).

\begin{figure}
  \centering 
  \includegraphics[trim={0cm 2cm 0cm 2cm},clip,width=0.75\textwidth]{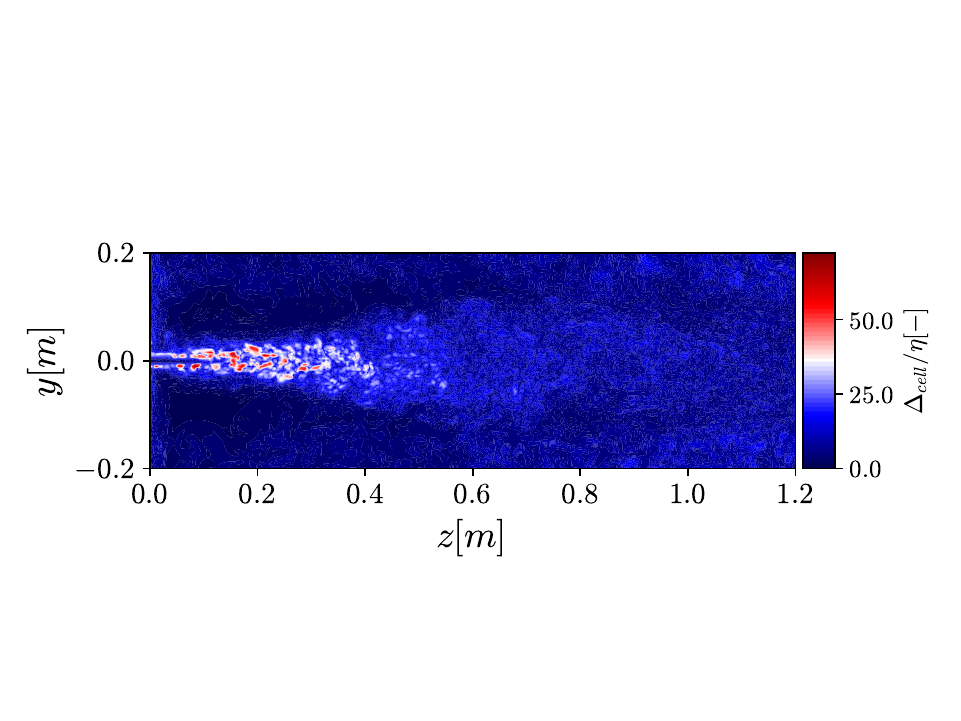}
  \caption{Meridional distribution of the instantaneous grid resolution relative to the local Kolmogorov length scale. The operating conditions are detailed in table \ref{tab:les_ambient_jet}.}
  \label{fig:grid_resolution}
\end{figure}

To assess the LES grid resolution, the ratio of the cell size $\Delta_{cell}$ (equal to the cubic root of the cell volume) to the local Kolmogorov length scale $\eta$ (determined from the resolved and modelled turbulent dissipation rate $\varepsilon$) is evaluated. For the operating conditions used in this work (see tables \ref{tab:les_ambient_jet} and \ref {tab:les_pbe_combined} below), $\Delta_{cell}$ is between approximately $(30-80)\eta$, see figure \ref{fig:grid_resolution}; this is typical for LES of shear flows \citep{You2007}. 

The PBE grid consists of 35 geometrically distributed bins (i.e.\ uniform in logarithmic space); this is  sufficient to capture the zero-th and first moments, and the water vapour consumption rate. The PBE grid resolution is assessed in appendix \ref{app:pbe_grid_refinement}. The initial aerosol particles have size below $100 \textnormal{ nm}$, the size of the first bin is $d_m=1.2\,\mathrm{nm}$. Particle sizes at the end of the computational domain remain of the order of a few microns. 

The LES simulations are initialised with the ambient conditions  and evolved in time for approximately 10 through-flow times (computed from the average jet centreline velocity and domain length). This is sufficient to recover the cylindrical symmetry of the flow, and achieve statistically steady state aerosol characteristics. For selected flow variables  $q(\textbf{x};t)$ statistics are computed on-the-fly for approximately $1.5$ through-flow times after the flow has reached statistical steady state. Statistics for the aerosol-phase are also computed on-the-fly. To increase the number of samples and improve statistical convergence, the cylindrical symmetry of the base flow is exploited and field variables are also azimuthally averaged. The time mean and the root-mean-square fluctuations of variable $q(\textbf{x};t)$ are denoted as $\overline{q}$ and $\sqrt{\overline{q^{\prime^2}}}$) respectively.

\section{Results and discussion}\label{sec:results}

\subsection{Validation of gas-phase results}\label{sec:results_les}
\begin{figure}
  \centering 
  \includegraphics[trim={0cm 0cm 0cm 0cm},clip,width=\textwidth]{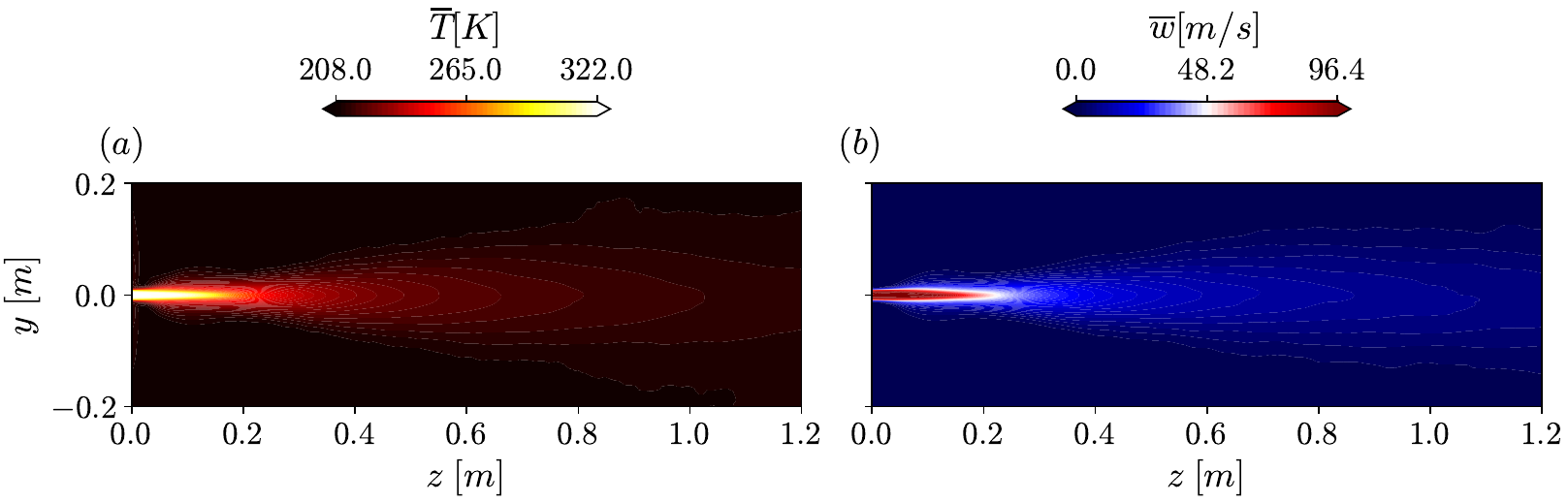}   
  \caption{Meridional distribution of time-averaged (\textit{a}) temperature and (\textit{b}) streamwise velocity (for the conditions of table \ref{tab:les_ambient_jet}).}
  \label{fig:axi_validation}
\end{figure}

\begin{figure}
  \centering 
  \includegraphics[trim={0cm 0cm 0cm 0cm},clip,width=\textwidth]{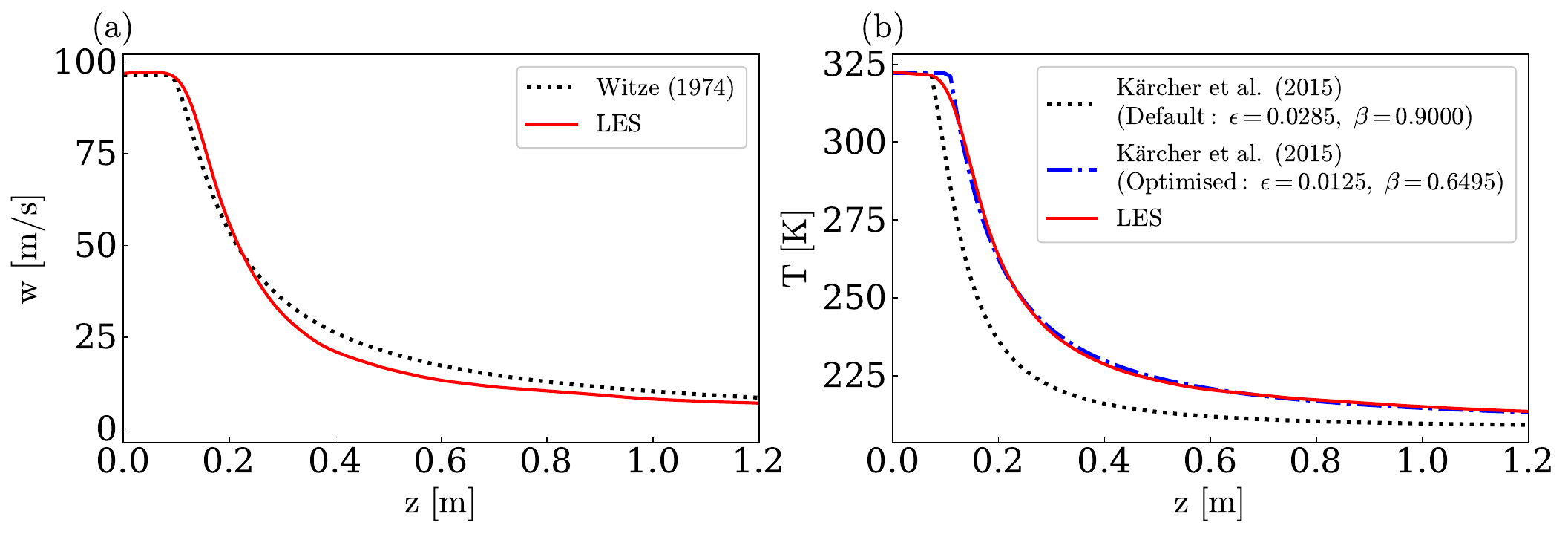}
  \caption{Comparison between LES data and empirical \citep{Witze1974} and semi-analytical \citep{Karcher2015} models of (\textit{a}) streamwise velocity and (\textit{b}) temperature profiles at the jet centreline.}
  \label{fig:centerline_models}
\end{figure}

To validate the LES model, we compare with the radial temperature profiles of \cite{Fagan2024}. The profiles were measured using thermocouples for a case without fuel injection. The ambient (rig) atmosphere and jet initial and boundary conditions are summarised in table \ref{tab:les_ambient_jet}. The hot jet flow is injected at $z=0$ in a cold and dry ambient environment. As it expands in the domain, it cools down (figure \ref{fig:axi_validation}a) and spreads radially (figure \ref{fig:axi_validation}b). The streamwise velocity ($w$) at the jet centreline is in overall  good agreement with the empirical model of \citet{Witze1974} for subsonic jets, refer to figure \ref{fig:centerline_models}a. For the temperature distribution, the semi-analytical model of \citet{Karcher2015} is used, 
\begin{equation}
    T(z)=T_a+\left(T_{j}-T_a\right)\left(\frac{\dfrac{r_{j}}{w_{j}}\sqrt{\dfrac{2}{\epsilon}}}{\max\!\left(t(z),\,\dfrac{r_{j}}{w_{j}}\sqrt{\dfrac{2}{\epsilon}}\right)}\right)^{\beta},
    \label{eq:temperature_K15}
\end{equation}
where $r_{j}$ is the jet-exit radius, $\beta$ is the dilution \citep{Karcher1999}, $\epsilon$ the  turbulent diffusivity \citep{Tollmien1926} and $t(z)$ is the time that it takes for the jet to travel from $0$ to $z$. The empirical model of \cite{Witze1974} for the axial velocity $w(z)$ is integrated to determine $t(z)$. With default parameters for $\beta$ and  $\epsilon$, the semi-analytical model predicts earlier break up compared to the LES data (figure \ref{fig:centerline_models}b). 
This is most likely because in the simulations turbulent fluctuations are not imposed at the jet exit, and this results in a delay of the jet break up. An optimisation of $\beta$ and $\epsilon$ to best fit the LES data indicates that the semi-analytical model is sufficiently flexible (figure \ref{fig:centerline_models}b). Compared to changes in the slope of the mixing line ($G^\mathrm{mix}$), changes in $\beta$ and $\epsilon$ have a secondary effect on the activation of the INPs and on the characteristics of the ice crystals at the end of the domain.

\begin{table}
  \begin{center}
\def~{\hphantom{0}}
\begin{tabular}{ccccccc}
\multicolumn{1}{c}{$p_{a/j}$ [Pa]} & 
\multicolumn{1}{c}{$T_a$ [K]} & 
\multicolumn{1}{c}{$T_j$ [K]} & 
\multicolumn{1}{c}{$w_a$ [m/s]} & 
\multicolumn{1}{c}{$w_j$ [m/s]} & 
\multicolumn{1}{c}{$Y_{O_2,a/j}$ [\%]} &
\multicolumn{1}{c}{$Y_{N_2,a/j}$ [\%]} \\[3pt] \hline
16235.7 & 
208  & 
322  &
0.0  &
96.4  &
$23.2$ &
$76.8$ 
\end{tabular}
\caption{Ambient ($a$) and jet exit ($j$) conditions for validation of gas-phase results.}
\label{tab:les_ambient_jet}
\end{center}
\end{table}

\begin{figure}
  \centering 
  \includegraphics[trim={0cm 0cm 0cm 0cm},clip,width=\textwidth]{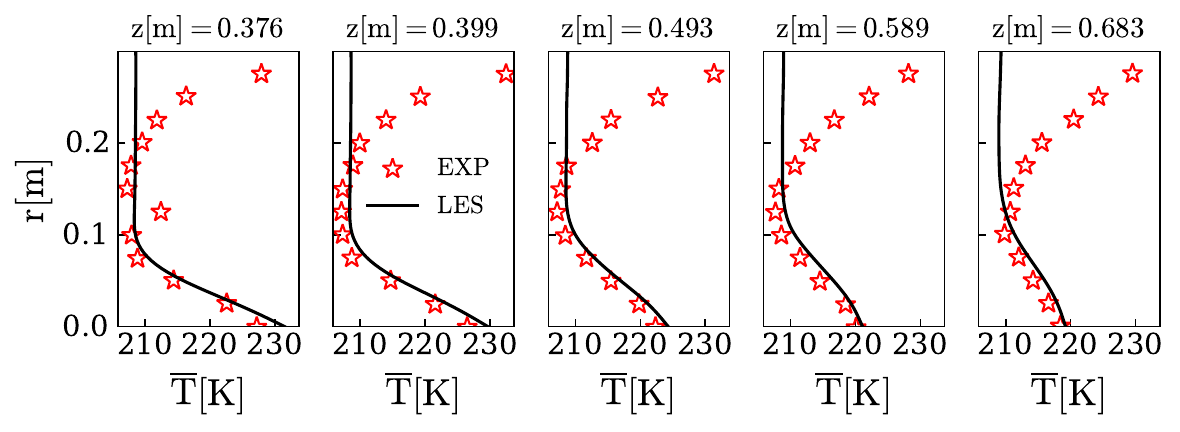}
  \caption{Temporally and azimuthally averaged radial profiles of mean static temperature at five streamwise (z) locations. Solid line: LES data; markers: experimental data (taken from figure 14 of \cite{Fagan2024}).}
  \label{fig:temperature_validation}
\end{figure}

Temporal- and azimuthally-averaged temperature distributions are compared with thermocouple measurements at 5 streamwise positions $z\in[0.35-0.7]m$ in figure \ref{fig:temperature_validation}. Close to the jet centreline ($r<0.1$), the LES matches the experimental data with a root-mean squared error (RMSE) below $\sim2.5\textnormal{K}$. Overall, this indicates that the LES can capture the jet-core cooling rate with sufficient accuracy. Far from the jet centreline ($r>0.1$), the agreement between LES and experiments deteriorates. In the experiments, the rig walls are not perfectly insulated from the surrounding environment at $T\approx288\textnormal{K}$. 
In the simulations however, due to lack of information about the wall temperature, the boundary is modelled as adiabatic, slip wall, see section \ref{sec:configuration}. One-dimensional jet modelling studies by \citet{Fagan2024} showed that the temperature distribution within the jet core is not significantly affected by the presence of the non-adiabatic walls. This is consistent with the present  LES results. 

\subsection{Ice particle characteristics and comparison with experimental data} \label{sec:results_les_pbe}

\begin{table}
\centering
\def~{\hphantom{0}}
\begin{tabular}{ccccccccccc}
& \multicolumn{6}{c}{Jet exit conditions ($j$)} & \multicolumn{4}{c}{Ambient conditions ($a$)} \\
\cline{2-7}\cline{8-11}
\begin{tabular}{@{}c@{}}Fuel\\(chem. formula)\end{tabular} &
\begin{tabular}{@{}c@{}}$T_j$\\$[$K$]$\end{tabular} &
\begin{tabular}{@{}c@{}}$w_j$\\$[$m\,s$^{-1}$$]$\end{tabular} &
\begin{tabular}{@{}c@{}}$Y_{H_2O,j}$\\$[\%]$\end{tabular} &
\begin{tabular}{@{}c@{}}$Y_{CO_2,j}$\\$[\%]$\end{tabular} &
\begin{tabular}{@{}c@{}}$Y_{O_2,j}$\\$[\%]$\end{tabular} &
\begin{tabular}{@{}c@{}}$Y_{N_2,j}$\\$[\%]$\end{tabular} &
\begin{tabular}{@{}c@{}}$p_a$\\$[$Pa$]$\end{tabular} &
\begin{tabular}{@{}c@{}}$T_a$\\$[$K$]$\end{tabular} &
\begin{tabular}{@{}c@{}}$w_a$\\$[$m\,s$^{-1}$$]$\end{tabular} &
\begin{tabular}{@{}c@{}}$Y_{N_2,a}$\\$[\%]$\end{tabular} \\[3pt]
\hline
\begin{tabular}{@{}c@{}}butanol\\($C_4H_{10}O$)\end{tabular} &
363 & 96.4 & 1.8 & 3.5 & 22.1 & 72.6 &
18754 & 225 & 1.0 & 1.0 \\
\begin{tabular}{@{}c@{}}H-S\\($C_{12}H_{26}$)\end{tabular} &
363 & 96.4 & 2.0 & 4.4 & 21.8 & 71.8 &
18754 & 225 & 1.0 & 1.0 \\
\end{tabular}
\caption{Jet exit ($j$) and ambient ($a$) conditions for contrail model assessment.}
\label{tab:les_pbe_combined}
\end{table}

Contrail characteristics from two different fuels are used for validation; a soot-rich fuel (n-butanol, referred to as $\textnormal{butanol}$) and SAF (HEFA-SPK, referred to as $\textnormal{H-S}$). The ambient conditions are summarised in table \ref{tab:les_pbe_combined}, and are the same for both fuels. A $w_a\approx0.01w_j$ co-flow is used, and the ambient gas is dry, gaseous nitrogen as in \citet{Fagan2024}. The choice of $\textnormal{butanol}$ as conventional fuel (instead of $\textnormal{Jet-A}$) is motivated by the unique surrogate chemical formula and the availability of fuel characteristics \citep{butanol_surrogate_pubchem} needed to determine the detailed initial gas composition. Experimental data were collected in a small volume $\Delta V =\Delta x \times\Delta y \times \Delta z \approx 0.025 \times 0.1 \times0.025 \, m^3$ centred at the jet centreline, approximately 64 jet diameters downstream of the notional exhaust $(x,y,z)=(0,0,1.15)m$. For consistency, the computational data were also averaged over the same volume before comparison with the experiments.

The number concentration of non-volatile (soot) particles $N_{nvPM}$ and their size characteristics were measured in the combustion chamber, before release in the cold, ambient rig atmosphere.  For $\textnormal{butanol}$, the experimental ice-number concentration ($N_{ice}$) at the measurement location $(x,y,z)=(0,0,1.15)$ m was lower than the number concentration of the non-volatile (soot) particles, $N_{ice}<N_{nvPM}$. This is an indication that the $\textnormal{nvPM}$ are mainly responsible for the ice crystal nucleation in the jet plume, which is expected for conventional, kerosene-based fuels \citep{Karcher2015}. On the other hand, for $\textnormal{H-S}$ the experimental ice-number concentration was greater than the injected $\textnormal{nvPM}$, $N_{ice}>N_{nvPM}$. As discussed in the introduction, in soot-poor regimes (typical for SAF) volatile particles ($N_{vPM}$) can activate, and lead to ice crystal formation under heterogeneous nucleation. However, the $\textnormal{vPM}$ properties at the exit of the combustion chamber were not provided in the experiments. 

The gas chemical composition is determined from the experimental equivalence ratio, surrogate formulas for n-butanol \citep{butanol_surrogate_pubchem} and HEFA-SPK \citep{hefa_spk_surrogate}, and assuming ideal (complete) combustion for oxygenated hydrocarbons \citep{Turns2000}. Table \ref{tab:inps_details} summarises the characteristics of the INPs at the jet exit ($r<d_j/2$, $z=0\textnormal{ m}$). For $\textnormal{butanol}$ only $\textnormal{nvPM}$ (soot) particles are used, while for $\textnormal{H-S}$ three cases are considered ($\textnormal{H-S-1}$, $\textnormal{H-S-2}$ and $\textnormal{H-S-3}$) with different initial number concentration of $\textnormal{vPM}$. All INPs are initially assumed to follow lognormal size distributions, characterised by the geometric mean diameter ($\textnormal{GMD}$) and the dimensionless geometric standard deviation ($\textnormal{GSD}$).

For the $\textnormal{butanol}$ and $\textnormal{H-S-1}$ cases, the initial INPs are all $\textnormal{nvPM}$ with number concentration ($N_{nvPM}$) equal to the experimental value, while $\textnormal{GMD}$, $\textnormal{GSD}$ and $\kappa$ are based on \cite{Ponsonby2025}. In the experiments, a $\textnormal{GMD}$ for the $\textnormal{nvPM}$ that formed in the combustion chamber is of the order of $\sim 5 \textnormal{ nm}$. However, ice-crystal nuclei are typically aggregates of soot primary particles \citep{Yu2024} with a typical GMD~$\approx35\textnormal{ nm}$ \citep{Karcher2015}; the latter is used to initialise the $\textnormal{nvPM}$ (table \ref{tab:inps_details}). For the $\textnormal{H-S-2}$ and $\textnormal{H-S-3}$  cases, the $\textnormal{nvPM}$ with a diameter $d_m<3\textnormal{ nm}$ are replaced with $\textnormal{vPM}$, with $N_{vPM}$, $\textnormal{GMD}$, $\textnormal{GSD}$ and $\kappa$ as specified in table \ref{tab:inps_details}. The $N_{vPM}$ was not provided in the experiments; it was determined based on the experimental fuel flow rate, and assuming HEFA-SPK fuel density $\rho_{H-S}\approx770 kg/m^3$ \citep{hefa_spk_density} and a typical emission index $EI_{vPM}\approx10^{15}-10^{17}$  \citep{Voigt2025_NATURE,Ponsonby2025}. Similarly, the values of $\textnormal{GMD}$, $\textnormal{GSD}$ and $\kappa$ for the $\textnormal{vPM}$ are based upon previous experimental and numerical studies \citep{Ponsonby2025}. The three cases for the HEFA-SPK fuel serve to determine the role of $\textnormal{vPM}$ activation on ice crystal formation, and to quantify $N_{ice}$ and $d_{mean,ice}$ sensitivities to changes in $N_{vPM}$, with the main difference between $\textnormal{H-S-2}$ and $\textnormal{H-S-3}$ being the initial concentration in $\textnormal{vPM}$.

\begin{table}
  \begin{center}
  \begin{tabular}{c c c}
    \multicolumn{1}{c}{Case} &
    \multicolumn{1}{c}{($N$, GMD, GSD, $\kappa)_{\textnormal{nvPM}}$} &
    \multicolumn{1}{c}{($N$, GMD, GSD, $\kappa)_{\textnormal{vPM}}$} \\[3pt]

    \hline
    
    Butanol &
    $(4.47\times10^{12},\, 35,\, 2.0,\, 0.005)$ &
    -- \\
    
    H-S-1 &
    $(1.05\times10^{11},\, 35,\, 2.0,\, 0.005)$ &
    -- \\
    
    H-S-2 &
    $(1.05\times10^{11},\, 35,\, 2.0,\, 0.005)$ &
    $(2.1\times10^{13},\, 2,\, 1.3,\, 0.5)$ \\
    
    H-S-3 &
    $(1.05\times10^{11},\, 35,\, 2.0,\, 0.005)$ &
    $(2.1\times10^{12},\, 2,\, 1.3,\, 0.5)$
    
  \end{tabular}
  \caption{Particle characteristics for nvPM (and vPM, if present) for the 4 cases considered. The units are $N$[\#/m$^3$], GMD[nm], GSD[-], $\kappa$[-].} 
  \label{tab:inps_details}
  \end{center}
\end{table}

\begin{figure}
  \centering 
  \subfloat[]{\includegraphics[trim={0cm 0cm 0cm 0cm},clip,width=0.5\textwidth]{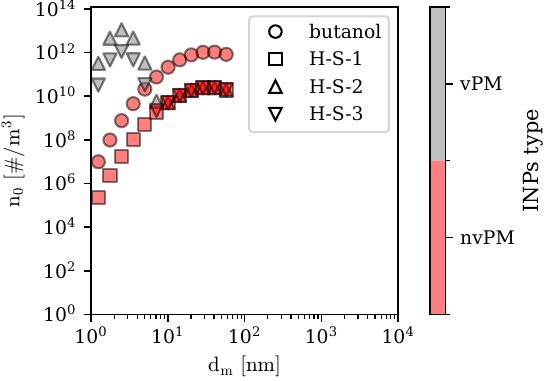}\label{fig:inps}} 
  \hfill 
  \subfloat[]{\includegraphics[trim={0cm 0cm 0cm 0cm},clip,width=0.5\textwidth]{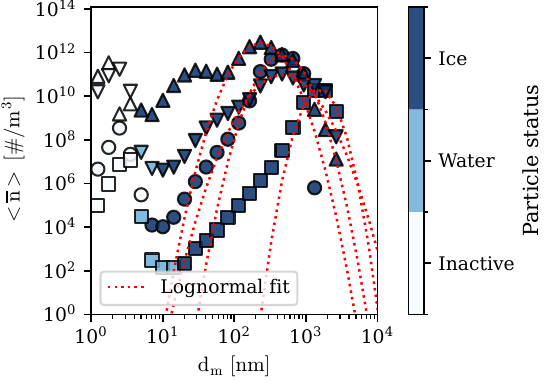}\label{fig:nv_validation}} 
  \caption{(\textit{a}) Initial number density distribution for the INPs for the 4 cases detailed in table \ref{tab:inps_details}; (\textit{b}) time- and volume-average particle number density distribution at $z=1.15 \textnormal{ m}$. In (\textit{b}) the dotted red lines indicate lognormal fits to the ice crystals size distributions.}
  \label{fig:pdf_validation}
\end{figure}

The initial INP diameter distributions for all 4 cases ($\textnormal{butanol}$ and 3 $\textnormal{H-S}$) are depicted in figure \ref{fig:inps}. As mentioned earlier, the data in the LES-PBE simulations are sampled at $(x,y,z)=(0,0,1.15)$ m and averaged over the same probe volume $\Delta V$ used in the experiments. The volume averaging operator is indicated with $<\cdot>$ and the time and volume averages aerosol distributions are depicted in figure \ref{fig:nv_validation}. The particle status (inactive, water droplet or ice particle) is also indicated. As can be seen, a significant portion of the smallest INPs remain inactive; this is because they require a higher saturation ratio ($S_v$) to activate, see equation \ref{eq:kappa_kohler} and figure  \ref{fig:kappa_kohler}. At the measurement location, the homogeneous freezing criterion for water droplets to turn into ice crystals (appendix \ref{app:ice_microphysics}) is satisfied for nearly all the active particles due to the combination of cold jet temperature and available liquid water volume \citep{Karcher2015}. A lognormal fit to the ice-crystal size distribution predicted by the discretised PBE approach suggests that the usual lognormal assumption \citep{Fagan2024,Lottermoser2025} becomes less valid when vPM are included among the INPs (H-S-2 and H-S-3 cases). Nevertheless, the coefficient of determination is high, $R^2>0.95$, and this could potentially be improved by using two lognormal distributions and separating ice crystals that originate from $\mathrm{nvPM}$ and $\mathrm{vPM}$, respectively.

\begin{figure}
  \centering 
  \includegraphics[trim={0cm 0cm 0cm 0cm},clip,width=1.0\textwidth]{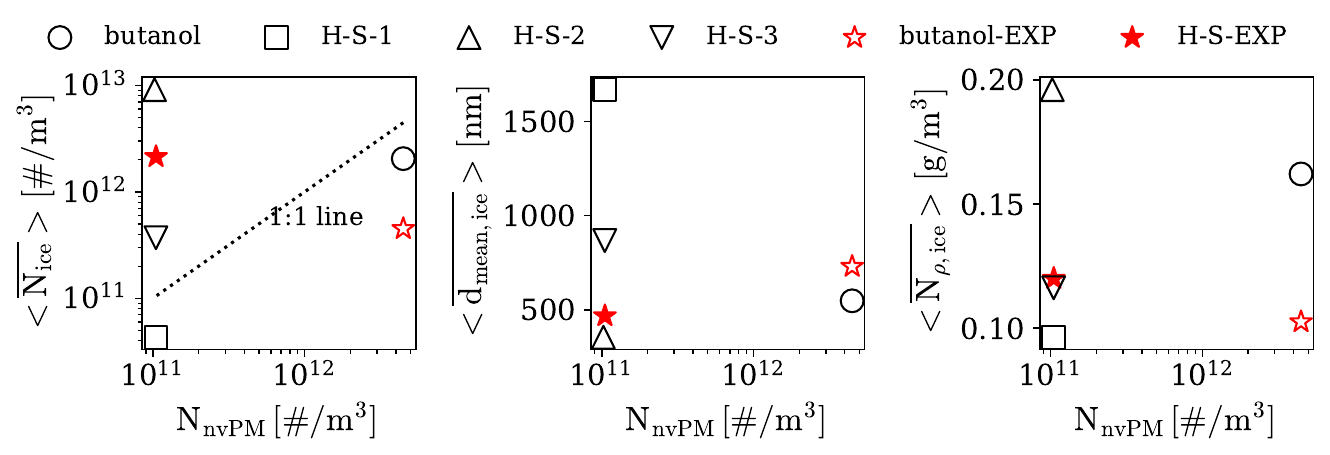}
  \caption{Comparison between the LES-PBE results (black) and experimental data (red) for the time- and volume-average ice crystal statistics. Left to right: ice crystal number concentration, mean diameter, and mass density. The rms fluctuation range for the LES-PBE data is approximately of the size of the marker. The experimental data can be found in figures 16-19 of \citet{Fagan2024}.}
  \label{fig:ice_statistics_validation}
\end{figure}

Time- and volume-average ice crystal statistics are shown in figure \ref{fig:ice_statistics_validation} for the four cases. For $\textnormal{butanol}$, $\sim90\%$ of the initial INPs activated into ice-crystals, and the predicted $N_{ice}$ is approximately 4.5 times larger compared to the measured data (left panel). The greater number of activated particles in the CFD compared to experiments lead to a nearly 25\% smaller ice mean diameter (mid panel) due to the larger consumption of water vapour saturation that prevents further growth in the numerical model relative to experiments. As a result of the larger ice crystal concentration in the simulations, the ice mass density ($N_{\rho,ice}$) is also overestimated by a factor of $\sim1.6$ (right panel).

For $\textnormal{H-S-1}$, only $\textnormal{nvPM}$ are introduced through the jet, and the resulting ice crystal concentration is nearly two orders of magnitude lower compared to experiments (left panel). The ice mean diameter is approximately $1770$~nm in the CFD, whereas it is $\sim730$~nm in the experiments, and the ice mass density is also slightly underestimated by the simulations. This indicates that the depletion of water vapour supersaturation ($s_v = S_v - 1$) is lower relative to the experiments, and this is a consequence of the lower predicted concentration of active aerosol particles. The introduction of $\textnormal{vPM}$ in the simulations (cases $\textnormal{H-S-2}$ and $\textnormal{H-S-3}$), which, together with $\textnormal{H-S-1}$, are compared against the single experiment $\textnormal{H-S}$ and differ only in the assumed $\textnormal{vPM}$, changes significantly the results relative to $\textnormal{H-S-1}$. The activation of $\textnormal{vPM}$ results in significantly higher ice crystal concentration and smaller ice mean diameters. While the lack of more detailed information regarding the presence and characteristics of $\textnormal{vPM}$ in the experiments prevents a more comprehensive model validation, it can be seen that by varying $N_{vPM}$ the LES-PBE model results can bracket the experimental data. 

In the following sections, the numerical model will be used to determine the role of flow inhomogeneity and turbulence on contrail formation from soot-rich fuels and SAF.

\subsection{Spatial distribution of ice particle characteristics} \label{sec:results_spatial_temporal_effect}

\subsubsection{Soot-rich fuel: $\textnormal{butanol}$}\label{subsec:butanol}
\begin{figure}
  \centering 
  \includegraphics[trim={0cm 0cm 0cm 0cm},clip,width=\textwidth]{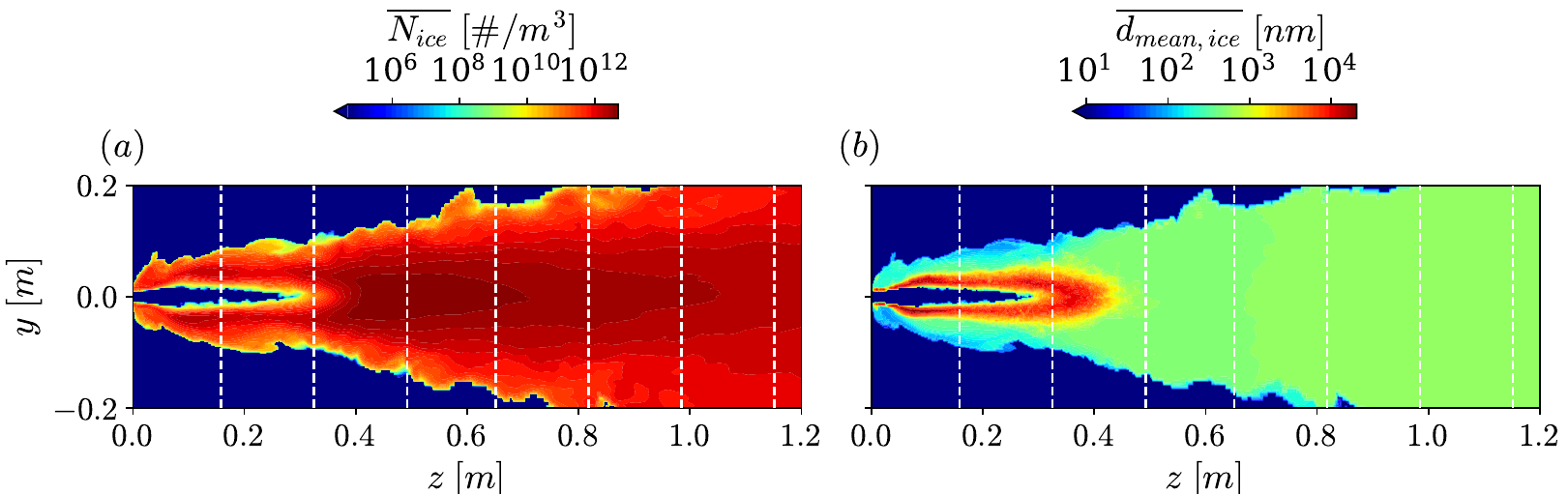}   
  \caption{Meridional distribution of time-averaged (\textit{a}) ice-number concentration and (\textit{b}), ice mean diameter for butanol case. The vertical dashed lines mark the integration locations for data shown in figure \ref{fig:ridge_plot_butanol}.}
  \label{fig:N_ice_D_ice_mean_butanol}
\end{figure}

Figure \ref{fig:N_ice_D_ice_mean_butanol}a shows contours of time-average ice-number concentration for $\textnormal{butanol}$. The initial activation of aerosol particles occurs in the jet shear layer ($|y|>0.5 d_j$) where ambient cold air is entrained and increases the supersaturation. This was also recently experimentally observed by \citet{Roy2026}. The ice-number concentration at the jet centreline starts to increase rapidly approximately at $(y,z)\approx(0,0.3)\,\textnormal{m}$, 17 jet diameters downstream of the jet exit. The kinetics of the ice particles is initially quite fast, driven by the high supersaturation relative to ice \citep{Karcher2015}, leading to a rapid increase in the ice mean diameter that locally can reach around $\overline{d_{mean,ice}}\approx10 \,\mu m$ at  $z\in[0.35,0.45]\,\textnormal{m}$, see figure \ref{fig:N_ice_D_ice_mean_butanol}b. 

\begin{figure}
  \centering 
  \includegraphics[trim={0cm 0cm 0cm 0cm},clip,width=\textwidth]{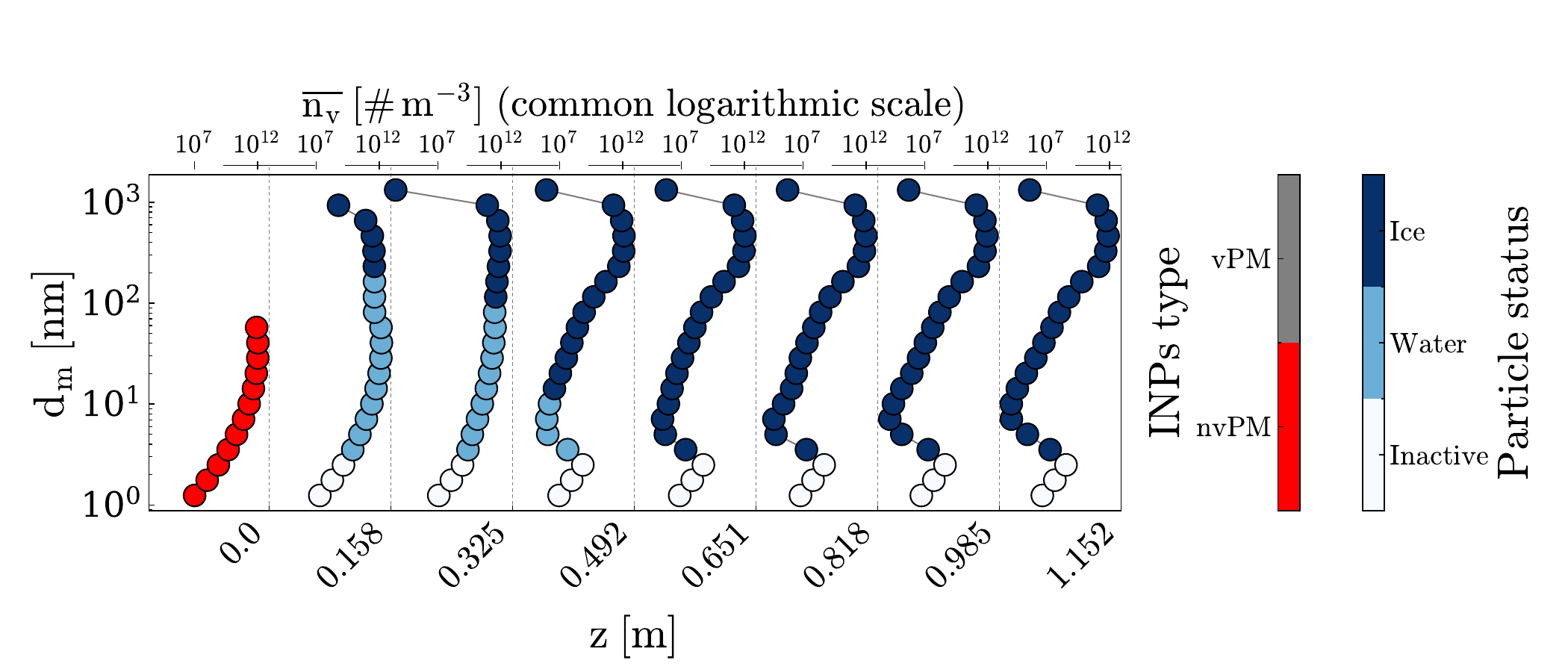}   
  \caption{Ridge plot of the weighted-average aerosol number concentration in the cross-section ($r\leq0.2\textnormal{ m}$) at various streamwise locations (marked by white dashed lines in figure \ref{fig:N_ice_D_ice_mean_butanol}) for case $\textnormal{butanol}$. The weighted averaging is based on the local mass-flow rate. Each marker represents a particle bin (note that the vertical axis is particle diameter). At $z=0\,\mathrm{m}$ the markers are coloured based on the INPs type (only nvPM for butanol), while for distributions at $z>0\,\mathrm{m}$ markers are coloured based on the particle status (inactive, water, ice), see  bar on the right.}
  \label{fig:ridge_plot_butanol}
\end{figure}

The growth of the ice mean diameter is driven by the activation of the largest INPs. This can be seen more clearly by inspecting the ridge plot in figure \ref{fig:ridge_plot_butanol}. The plot shows the number concentration and the status of each bin at the streamwise locations marked by the white dashed lines in figure  \ref{fig:N_ice_D_ice_mean_butanol}. At each location, the cross-section data are weighted-averaged using the local mass-flow rate. Closer to the jet exit ($z\lesssim0.5\ \textnormal{m}$ in figure \ref{fig:ridge_plot_butanol}), the largest aerosols activate first into supercooled water droplets and condensational growth starts. The largest droplets freeze first due to the greater liquid water volume, $\textnormal{LWV}$, that reduces the freezing temperature (see section \ref{app:freezing_criterion} for details) and lead to a large ice mean diameter with  a peak at $z\approx0.5\ \textnormal{m}$ (figure \ref{fig:N_ice_D_ice_mean_butanol}b). 
As the smallest particles also activate in larger concentrations and homogeneously freeze forming ice crystals, the ratio of the first moment to the zeroth moment (equation \eqref{eq:d_ice_mean}) reduces. Thus, the ice mean diameter reduces, and the water vapour supersaturation is consumed. The kinetics slows down, the ice-number concentration reduces due to dilution and the size of the ice crystal remains quite uniform in space, with $\overline{d_{mean,ice}}\approx1 \,\mu m$, as shown in figure \ref{fig:N_ice_D_ice_mean_butanol}. 

\begin{figure}
  \centering 
  \includegraphics[trim={0cm 0cm 0cm 0cm},clip,width=\textwidth]{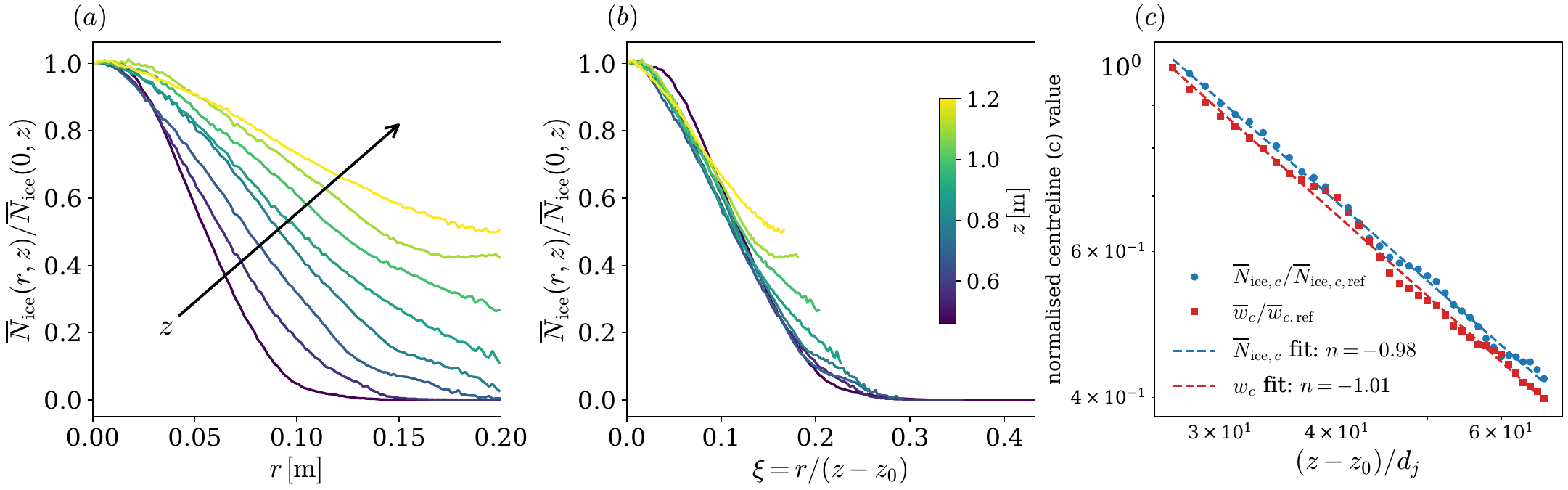} 
\caption{Self-similarity and downstream decay of the ice field for the \textnormal{butanol} case. (a) Normalised radial profiles of ice-number concentration 
$N_{\mathrm{ice}}(r,z)/N_{\mathrm{ice}}(0,z)$, at selected downstream locations. (b) Normalised profiles plotted against the classical self-similarity coordinate $\xi=r/(z-z_0)$, where $z_0$ is the virtual origin. (c) Centreline ice-number concentration and mean axial velocity downstream of the location of maximum value of ice-number concentration.}
  \label{fig:self_similarity_butanol}
\end{figure}

The time- and azimuthally averaged ice-number-concentration profiles  across the jet are plotted in figure \ref{fig:self_similarity_butanol} for $z\gtrsim0.4\textnormal{ m}$, i.e.\ downstream of the peak, see figure \ref{fig:N_ice_D_ice_mean_butanol}b. At each $z$ location, the radial profiles are normalised by the centreline value, conventionally denoted by $N_{\mathrm{ice}}(0,z)$. When plotted against the radial coordinate $r$, the  profiles become flatter due to the the spreading of the jet and the better mixing, see figure \ref{fig:self_similarity_butanol}a.  Following the conventional self-similar description of scalar transport in a spreading round jet \citep{Nguyen2025}, the profiles are also plotted against the dimensionless coordinate $\xi=r/(z-z_0)$, where $z_0$ is the virtual origin of the developed jet, here taken as $z_0=0$ (the exit plane). The profiles $N_{\mathrm{ice}}(r,z)/N_{\mathrm{ice}}(0,z)$ almost collapse in a single curve, as shown in figure \ref{fig:self_similarity_butanol}b. This indicates that the radial ice-number-concentration profile is approximately self-similar in the developed region. The centreline ice-number concentration and mean axial velocity are plotted against $\xi$ in figure \ref{fig:self_similarity_butanol}c. The decay of both quantities is well captured by a power law of the form
\begin{equation}
\overline{N}_{\mathrm{ice}}(0,z) \propto (z-z_0)^{n_N},
\qquad
\overline{w}(0,z) \propto (z-z_0)^{n_w}.
\end{equation}
The exponents $n_N$ and $n_w$ are obtained by  least-squares regression. Taking the location of maximum centreline ice-number concentration $z_{\mathrm{ref}}$ as the reference position,  figure~\ref{fig:self_similarity_butanol}c shows that the power-law exponents $n_N$ and $n_w$ are close to $-1$. The nearly identical decay rates imply that $\overline{N}_{\mathrm{ice}}(0,z)/\overline{w}(0,z)$ remains approximately constant, which indicates that the decay of ice-number concentration is primarily governed by dilution and entrainment. This result 
indicates that a one-dimensional steady model may be sufficient to represent ice-number-concentration dilution under soot-rich conditions.

\begin{figure}
  \centering 
  \includegraphics[trim={0cm 0cm 0cm 0cm},clip,width=0.6\textwidth]{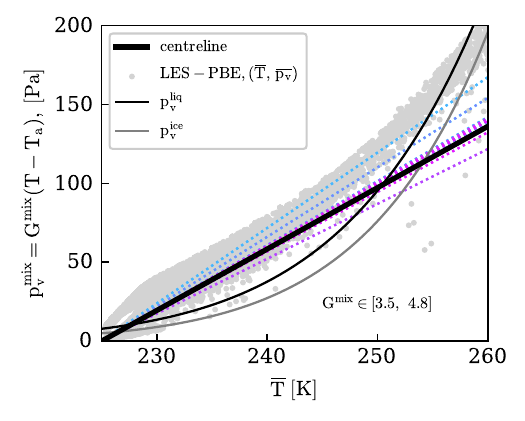}
  \caption{Mixing lines evaluated along the jet centreline (thick black solid line) and other flow trajectories (coloured dotted lines) for case $\textnormal{butanol}$ determined from the LES-PBE data as detailed in section \ref{sec:1d_pbe_method}. The grey and black curves indicate the ice and liquid vapour saturation, respectively. The light grey markers are  $\mathrm{(\overline{T},\overline{p_{\mathrm{v}}}})$ data along the flow trajectories from the LES-PBE simulations.}
  \label{fig:pvT_butanol}
\end{figure}

The mixing lines that satisfy the Schmidt-Appleman criterion are determined from the LES-PBE simulations as described in section \ref{sec:1d_pbe_method}. They were computed along various streamlines, including the centreline, and are plotted in figure \ref{fig:pvT_butanol}. Data points $\mathrm{(\overline{T},\overline{p_{\mathrm{v}}}})$ evaluated along these trajectories are also superimposed. These points account for the gas-aerosol thermodynamic feedback. At a fixed temperature, the data exhibit substantial spread due to radial non-uniformity across the jet, and only a few points approach the saturation curves. This occurs because the smallest INPs remain inactive ($r_d \lesssim 4\,\mathrm{nm}$ in figure \ref{fig:ridge_plot_butanol}) and water-vapour consumption by the active particles is insufficient to deplete the supersaturation.

\begin{figure}
  \centering 
  \subfloat[]{\includegraphics[trim={0cm 0cm 0cm 0cm},clip,width=0.5\textwidth]{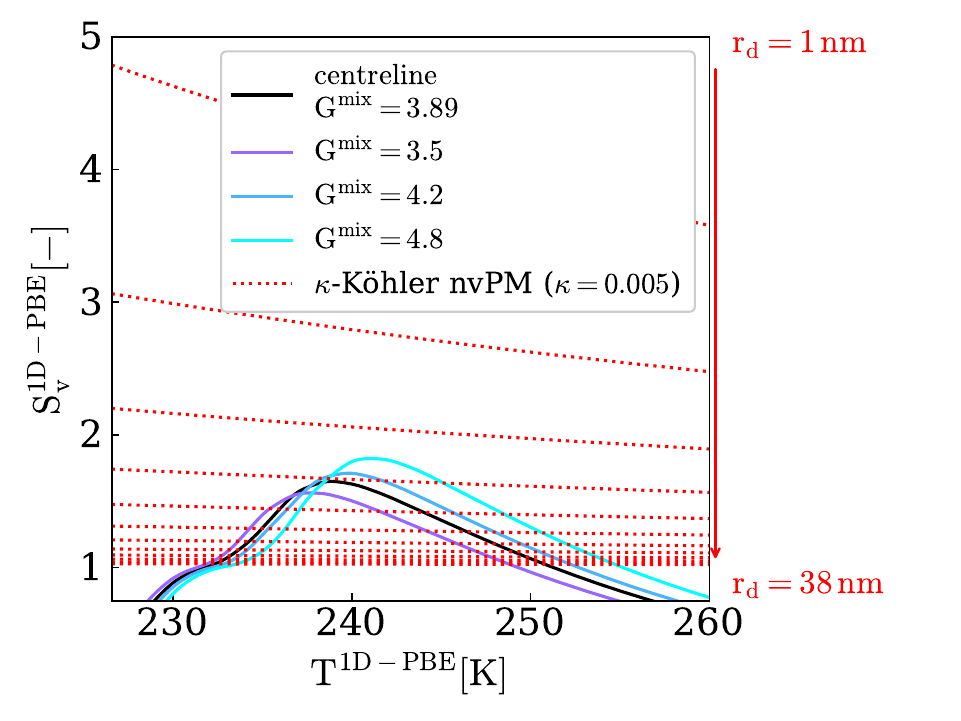}\label{fig:saturation_1d_pbe_butanol}}
  \hfill
  \subfloat[]{\includegraphics[trim={0cm 0cm 0cm 0cm},clip,width=0.5\textwidth]{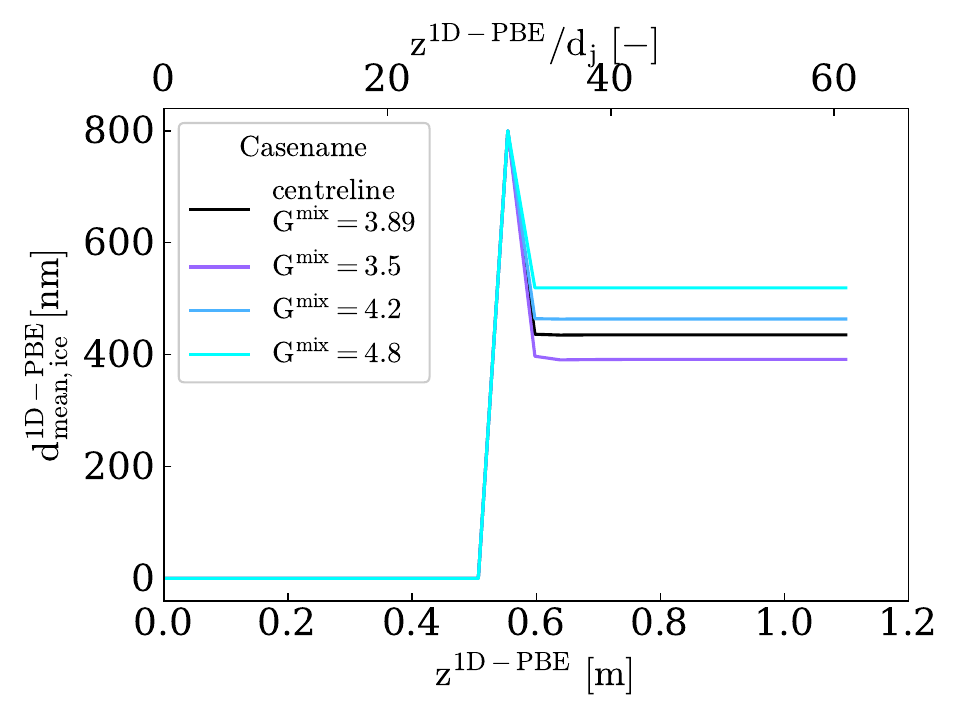}\label{fig:ice_mean_diam_1d_pbe_butanol}}
  \caption{1D-PBE results for $\textnormal{butanol}$. (\textit{a}) Liquid-water vapour saturation ratio as a function of the evolving jet/plume temperature $T^{\mathrm{1D-PBE}}$ for different values of $G^{\mathrm{mix}}$. The red-dotted curves mark the critical saturation ratio for each injected dry-particle diameter based on $\kappa$-Köhler.  (\textit{b}) Ice mean diameter for $\textnormal{butanol}$ versus the Witze-based spatial coordinate $z^{\mathrm{1D-PBE}}$.}
  \label{fig:1D_PBE_butanol}
\end{figure}

The values of $G^{\mathrm{mix}}$ obtained from the LES-PBE data are used as input for the 1D-PBE model with the same initial characteristics of the aerosol particles (case $\textnormal{butanol}$ in table \ref{tab:inps_details}). The liquid-water saturation ratio against the jet-plume temperature $T^{\mathrm{1D-PBE}}$ for different $G^{\mathrm{mix}}$ values is shown in figure \ref{fig:saturation_1d_pbe_butanol}.  The horizontal axis is the evolving temperature $T^{\mathrm{1D-PBE}}$ of the jet/plume 
and it is based on equation \eqref{eq:temperature_K15}. The initial value is the jet exit condition $T_{j}$. It can be seen that an increase in $G^{\mathrm{mix}}$ only slightly increases the temperature at which depletion of the water vapour saturation becomes significant, which is mainly a consequence of the activation of a larger fraction of the smallest injected $\textnormal{nvPM}$. At the end of the domain, when nearly all the supersaturation is consumed, the difference in the fraction of activated $\textnormal{nvPM}$ due to a change in $G^{\mathrm{mix}}$, has only slightly affected the ice mean diameter that remains bounded between $d_{mean,ice}^{\mathrm{1D-PBE}}=400\textnormal{ nm}$ and $550\textnormal{ nm}$, see figure \ref{fig:ice_mean_diam_1d_pbe_butanol}.  
In this figure, the streamwise coordinate $z^{\mathrm{1D-PBE}}$ is obtained from the elapsed 1D-PBE time $t$ for each parcel and the \cite{Witze1974} velocity model $w(z)$ by integrating $dz/dt=w(z)$. The $w(z)$ model is parameterised using the operating conditions listed in table \ref{tab:les_pbe_combined}. The sharp changes in the the profiles of the ice mean diameter are due to the instantaneous freezing model (see appendix \ref{app:freezing_criterion} for details). In agreement with figure \ref{fig:N_ice_D_ice_mean_butanol} for the LES-PBE data, the 1D-PBE model predicts a peak in the ice mean diameter at around $z^{\mathrm{1D-PBE}}\approx0.5 \textnormal{ m}$. This is because the largest aerosols activate and freeze first while the smallest INPs require a higher supersaturation to activate (see red-dotted curves in figure \ref{fig:saturation_1d_pbe_butanol}). Once these INPs also reach the homogeneous freezing conditions, the ice mean diameter decreases for $z^{\mathrm{1D-PBE}}\gtrsim0.6 \textnormal{ m}$. At the measurement location $z^{\mathrm{1D-PBE}}=1.15\textnormal{ m}$, the 1D-PBE model at the centreline and the LES-PBE data (shown in figure \ref{fig:ice_statistics_validation}) provide a similar estimate of the ice mean diameter ($\approx450-500\textnormal{ nm}$). On the other hand, the mean diameter for the whole aerosol ($d_{mean,aerosol}$) evolves smoothly as particles activate and  supercooled water droplets and ice crystals grow (see figure \ref{fig:D_grid_ref} in appendix \ref{app:pbe_grid_refinement}). 

Overall, a steady, 1D model tuned with LES dilution data may be sufficient to reliably predict ice crystal formation for soot-rich fuels, like butanol. This is in qualitative agreement with previous work \citep{Karcher2015,Lewellen2020,Ponsonby2025}. 

\subsubsection{Sustainable aviation fuel: $\textnormal{H-S}$}\label{subsec:SAF}
\begin{figure}
  \centering 
    \includegraphics[trim={0cm 0cm 0cm 0cm},clip,width=\textwidth]{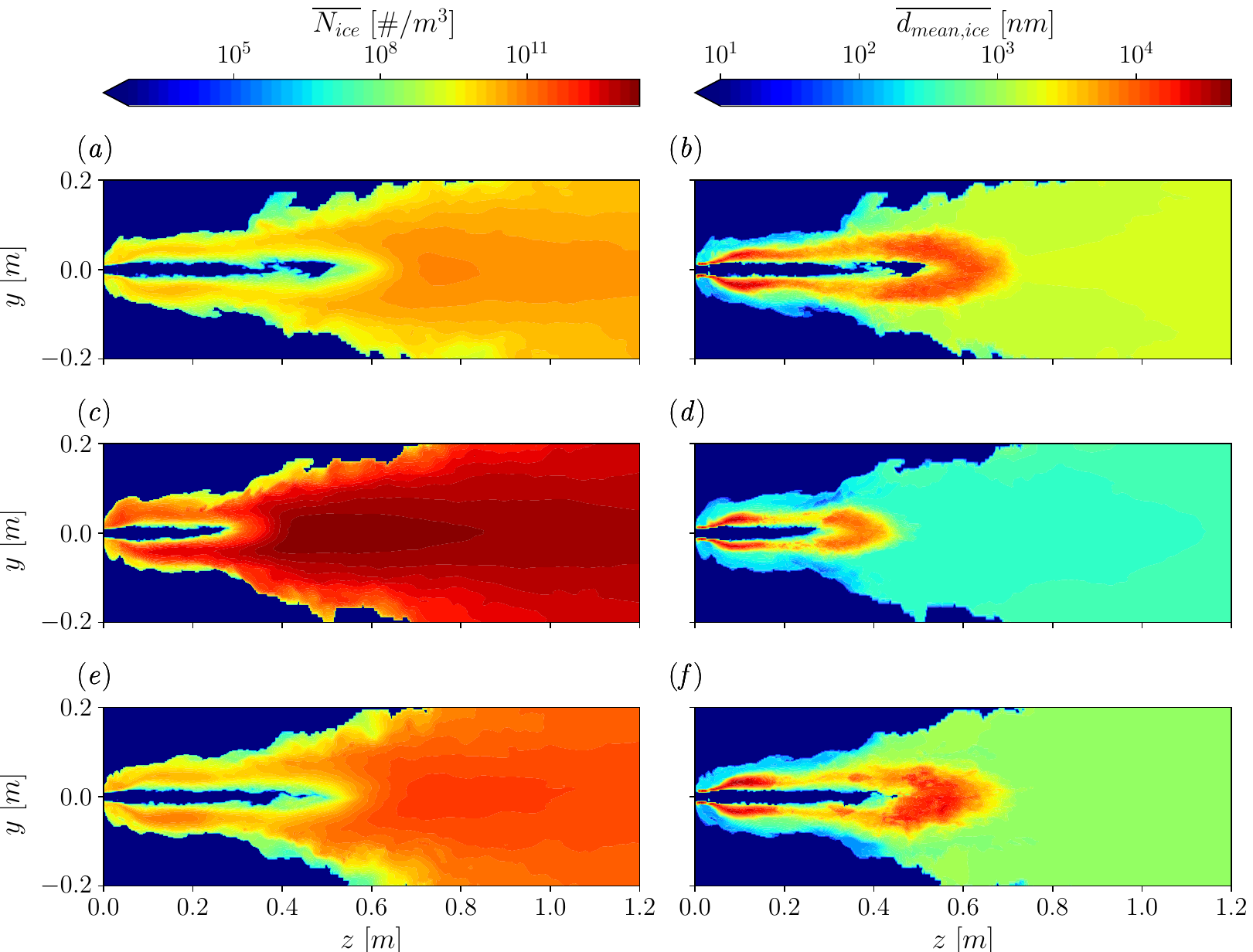}
  \caption{Meridional distribution of time-averaged (left column) ice-number concentration, and (right column) ice mean diameter. (\textit{a}), (\textit{b}): case H-S-1; (\textit{c}), (\textit{d}): case H-S-2; (\textit{e}), (\textit{f}): case H-S-3.}
  \label{fig:N_ice_D_ice_mean_hs}
\end{figure}

For SAF (HEFA-SPK) three cases are investigated, $\textnormal{H-S-1}$, $\textnormal{H-S-2}$ and $\textnormal{H-S-3}$, see table \ref{tab:inps_details}. They share the same jet exit and ambient gas conditions (table \ref{tab:les_pbe_combined}), but have different  $\textnormal{vPM}$ concentrations. The effect of   $\textnormal{vPM}$ activation on the ice crystal characteristics is depicted in figure \ref{fig:N_ice_D_ice_mean_hs}. The region of the initial formation of ice crystals remains the same; it is the shear layer with $|y|>r_j$ and it is not significantly affected by the presence of $\textnormal{vPM}$. However, significant differences exist in the core of the jet and in the  downstream evolution. More specifically, the presence of $\textnormal{vPM}$ in $\textnormal{H-S-2}$ and $\textnormal{H-S-3}$ (mid and bottom rows) leads to an earlier formation of ice crystals close to the jet centreline ($y=0$) compared to $\textnormal{H-S-1}$, at approximately 17 ($z\approx0.3$) and 23 ($z\approx0.4$) jet diameters downstream of the jet exit, respectively. Closer to the domain outlet ($z>0.8$ m), the activation of $\textnormal{vPM}$ leads to larger ice crystal number concentrations, compare panels (c) and (e) with panel (a), and smaller ice mean diameters, compare panels (d) and (f) with (b); this is driven by the larger consumption of water vapour supersaturation.

\begin{figure}
  \centering 
  \subfloat[]{\includegraphics[trim={0cm 0cm 0cm 0cm},clip,width=0.5\textwidth]{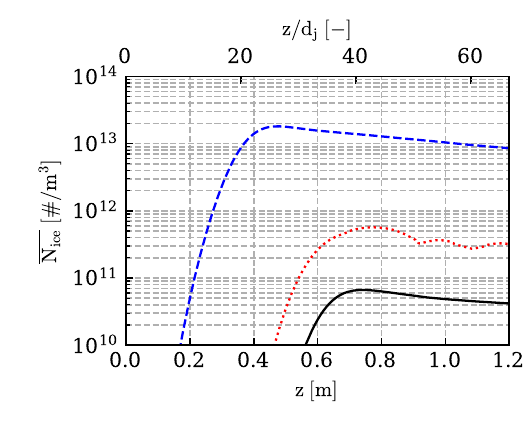}\label{fig:N_mean_avg_hs}} 
  \hfill 
  \subfloat[]{\includegraphics[trim={0cm 0cm 0cm 0cm},clip,width=0.5\textwidth]{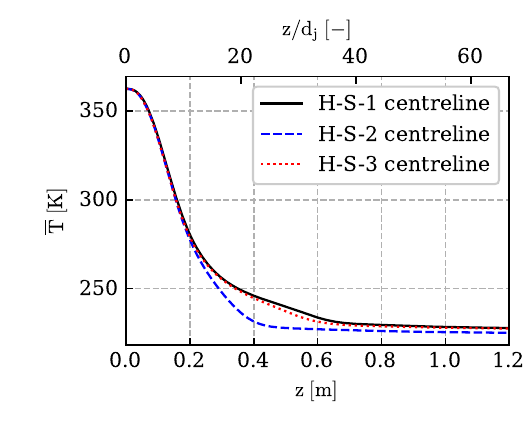}\label{fig:T_mean_avg_hs}}   
  \hfill
  \subfloat[]{\includegraphics[trim={0cm 0cm 0cm 0cm},clip,width=0.5\textwidth]{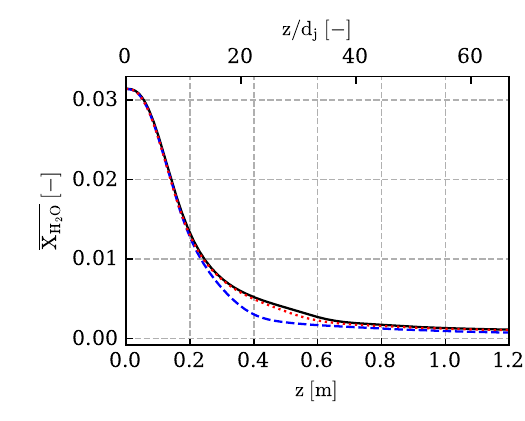}\label{fig:XH2O_mean_avg_hs}} 
  \hfill 
  \subfloat[]{\includegraphics[trim={0cm 0cm 0cm 0cm},clip,width=0.5\textwidth]{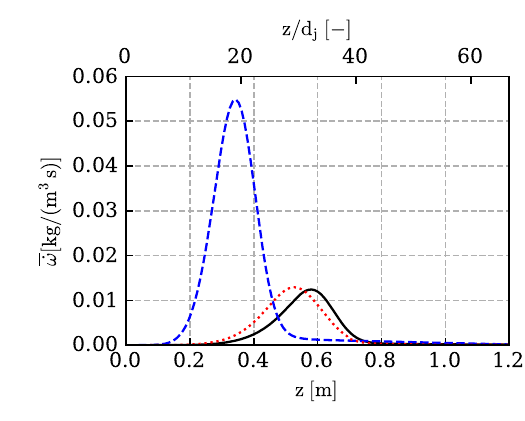}\label{fig:msource_mean_avg_hs}}      
  \caption{LES-PBE model of the effect of volatile particles on time-averaged, jet centreline distribution of (\textit{a}) ice-number concentration, (\textit{b}) static temperature, (\textit{c}) water vapour molar fraction, and (\textit{d}) water vapour consumption rate.}
  \label{fig:spatial_effect_hs}
\end{figure}

The presence and activation of $\textnormal{vPM}$ significantly affects the formation of ice crystals, see figure \ref{fig:N_mean_avg_hs}. For $\textnormal{H-S-3}$ the ice-number concentration is shifted upstream by approximately $5d_j$ compared to  $\textnormal{H-S-1}$. A further increase by one order of magnitude in the initial concentration of $\textnormal{vPM}$ (case $\textnormal{H-S-2}$ compared to $\textnormal{H-S-3}$) further shifts the profile upstream by about $15d_j$.  The peak values are also significantly affected;  $\max(\overline{N_{ice}}) \approx 6\times10^{11}\ \#/m^3$ and $\approx2\times10^{13}\ \#/m^3$for $\textnormal{H-S-3}$ and $\textnormal{H-S-2}$, respectively. These results are due to the non-linear interaction and dynamic coupling between the gas and aerosol phases; this coupling is further investigated in section \ref{sec:fluid_dynamics_activation}. 

Figure \ref{fig:T_mean_avg_hs} shows that as the number of $N_{vPM}$ is increased ($\textnormal{H-S-1}$, to $\textnormal{H-S-3}$, to $\textnormal{H-S-2}$) the jet temperature reduces to ambient temperature faster. This increase in jet cooling rate driven by the activation of $\textnormal{vPM}$ is probably due to the faster decay of water vapour molar fraction in the jet (see figure \ref{fig:XH2O_mean_avg_hs}), which changes the thermal conductivity of the mixture, and this is not compensated by the latent heat release due to phase change (from condensation and deposition). The initial kinetics of the $\textnormal{vPM}$ is fast \citep{Ponsonby2025}, driven by the small critical radii and large critical supersaturation for activation, leading to a notable increase (and upstream shift) in water vapour consumption rate for case $\textnormal{H-S-2}$ compared to $\textnormal{H-S-3}$ and $\textnormal{H-S-1}$, refer to figure \ref{fig:msource_mean_avg_hs}. As the consumption of water vapour supersaturation is faster for $\textnormal{H-S-2}$ relative to $\textnormal{H-S-3}$ and $\textnormal{H-S-1}$, the ice-number concentration $\overline{N_{ice}}$ reaches a peak further upstream ($z\approx0.45$) (and then decreases due to dilution). The results of figure \ref{fig:spatial_effect_hs} indicate strong, non-linear interaction among gas-phase and ice crystal growth, which is driven by the initial concentration and the microphysics of $\textnormal{vPM}$. 

\begin{figure}
  \centering
  \subfloat[]{\includegraphics[trim={0cm 0cm 0cm 0cm},clip,width=0.5\textwidth]{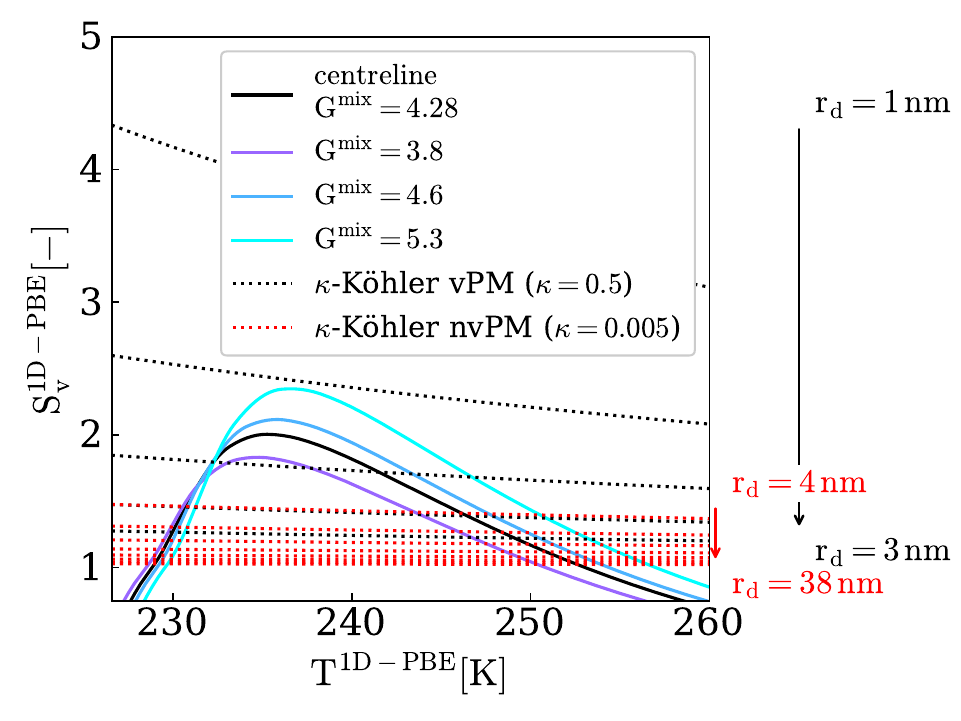}\label{fig:saturation_1d_pbe_hs3}}
  \hfill
  \subfloat[]{\includegraphics[trim={0cm 0cm 0cm 0cm},clip,width=0.5\textwidth]{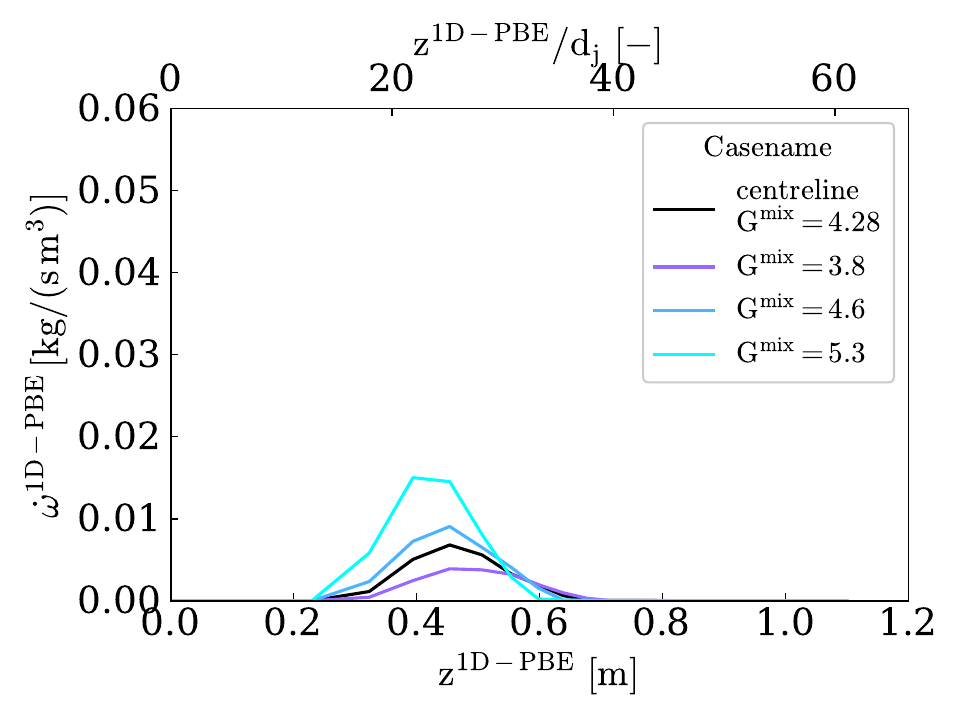}\label{fig:msource_avg_1d_pbe_hs3}}
  \hfill
  \subfloat[]{\includegraphics[trim={0cm 0cm 0cm 0cm},clip,width=0.5\textwidth]{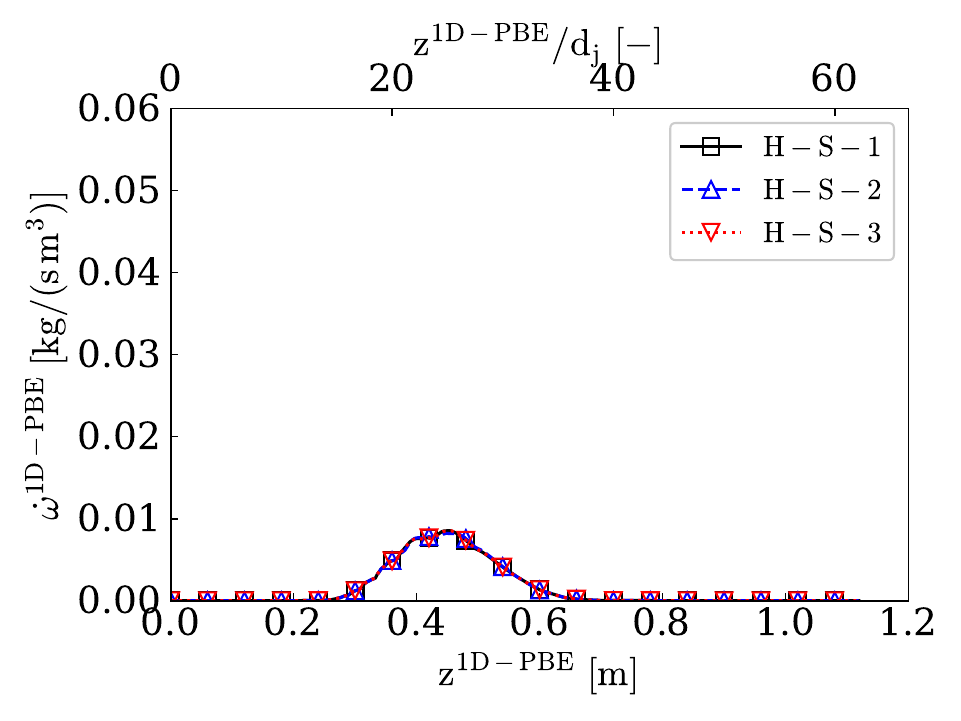}\label{fig:msource_avg_1d_pbe_hs}}
  \caption{(\textit{a}), (\textit{b}) 1D-PBE results for $\textnormal{H-S-3}$. (\textit{a}) Liquid-water vapour saturation ratio as a function of the evolving jet/plume temperature $T^{\mathrm{1D-PBE}}$ for different values of $G^{\mathrm{mix}}$, determined from the LES-PBE streamlines that satisfy the SAC. The coloured $\kappa$-Köhler curves give the critical saturation ratio for each injected dry-particle diameter. (\textit{b}) Volumetric water-vapour consumption rate along $z^{\mathrm{1D-PBE}}$. (\textit{c}) Mean 1D-PBE water-vapour consumption rate for $\textnormal{H-S-1}$, $\textnormal{H-S-2}$, and $\textnormal{H-S-3}$, averaged over the values of $G^{\mathrm{mix}}$.}
  \label{fig:1d-pbe-hs3-results}
\end{figure}

The 1D-PBE model is used to examine how the thermodynamic history of the jet affects particle activation for SAF. For each $\textnormal{H-S}$ case, the mixing-line slope $G^{\mathrm{mix}}$ is obtained from the LES-PBE data using $\mathrm{CO_2}$ as a passive scalar, as described in section \ref{sec:1d_pbe_method}. The values considered here span the streamlines satisfying the SAC, with $G^{\mathrm{mix}}\in[3.8,\,5.3]$. Figure \ref{fig:saturation_1d_pbe_hs3} for $\textnormal{H-S-3}$ is the equivalent of figure \ref{fig:saturation_1d_pbe_butanol} for butanol. As can be seen, increasing $G^{\mathrm{mix}}$ increases the peak liquid-water vapour saturation ratio. The largest critical saturation ratios for the injected dry particles in figure \ref{fig:saturation_1d_pbe_hs3} correspond to the vPM population (with $\kappa=0.5$, see dotted black lines), which is now present. An increase in $G^{\mathrm{mix}}$ affects the activation temperature of the INPs, allowing smaller dry aerosols to activate at warmer temperatures, but does not lead to the activation of smaller vPM particles. Consequently, this produces only a relatively narrow spread in the volumetric water-vapour consumption rate for $\textnormal{H-S-3}$ as a function of $z^{\mathrm{1D-PBE}}$, see figure \ref{fig:msource_avg_1d_pbe_hs3}. Finally, figure \ref{fig:msource_avg_1d_pbe_hs} compares this rate across $\textnormal{H-S-1}$, $\textnormal{H-S-2}$, and $\textnormal{H-S-3}$. For each case, the plotted curve is the arithmetic mean of the 1D-PBE consumption-rate profiles from the four represented streamlines, associated with the range of $G^{\mathrm{mix}}$ shown in figure \ref{fig:saturation_1d_pbe_hs3}. The results are identical. This  highlights a limitation of the 1D-PBE representation: even with the dilution parameter tuned using the mean LES-PBE flow field, it cannot reproduce the vPM activation and water vapour consumption rates predicted by the coupled LES-PBE model (figure \ref{fig:msource_mean_avg_hs}). The scale-resolved turbulent flow field produces local thermodynamic histories that cannot be represented by the mean-flow 1D-PBE model. 

The limitation motivates a deeper investigation of the dynamic interaction between the gas-phase and the mechanisms driving $\textnormal{vPM}$ activation and ice-crystal growth. This is the subject of sections \ref{sec:fluid_dynamics_activation} and \ref{sec:fluid_dynamics_crystal_growth}, respectively.

\subsection{Effect of turbulent fluctuations on $\textnormal{vPM}$ activation}\label{sec:fluid_dynamics_activation}
\begin{figure}
  \centering 
  \includegraphics[trim={0cm 0cm 0cm 0cm},clip,width=\textwidth]{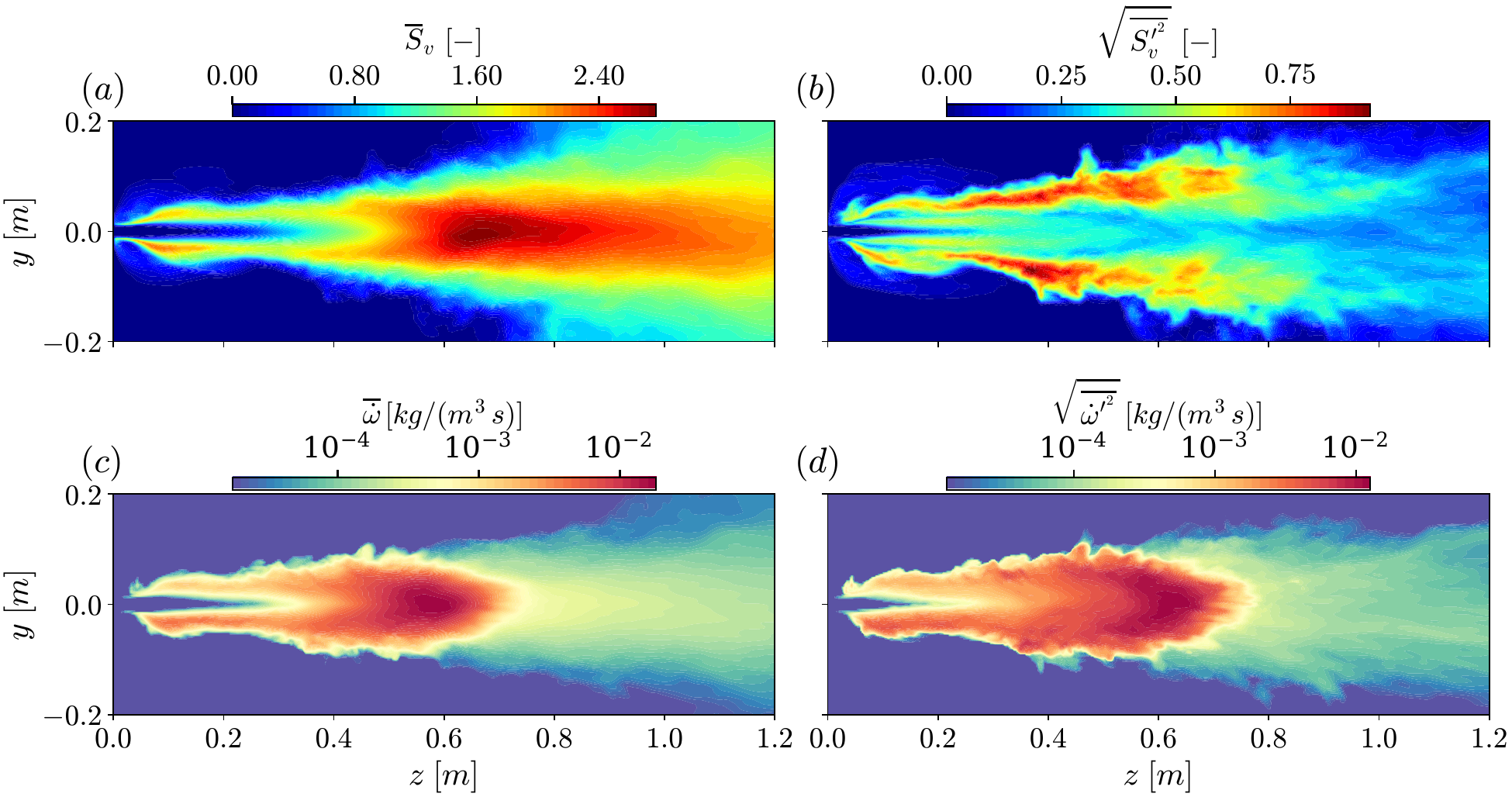}
  \caption{Meridional distribution of (left column) mean and (right column) rms fluctuation of (top row) saturation ratio and (bottom row) water vapour consumption rate for $\textnormal{H-S-3}$.}
  \label{fig:hs3_statistics}
\end{figure}

Figure \ref{fig:hs3_statistics}a shows contours of the the time-averaged saturation ratio $S_v$ for $\textnormal{H-S-3}$. The peak time-average value $\overline{S_v}\approx2.5$ is located close to the jet centreline. However, $S_v$ is highly unsteady with rms fluctuations $\sqrt{\overline{S_v^{\prime^2}}}\approx0.8$ closer to the jet shear layer (figure \ref{fig:hs3_statistics}b), where turbulent kinetic energy is produced due to the very high shear. Mean and rms fluctuations of water vapour consumption rate (figures \ref{fig:hs3_statistics}c and \ref{fig:hs3_statistics}d) both peak within the core of the jet plume, highlighting the role of turbulent mixing which enhances local supersaturation and drives aerosol activation and growth. 

\begin{figure}
  \centering 
  \includegraphics[trim={0cm 0cm 0cm 0cm},clip,width=\textwidth]{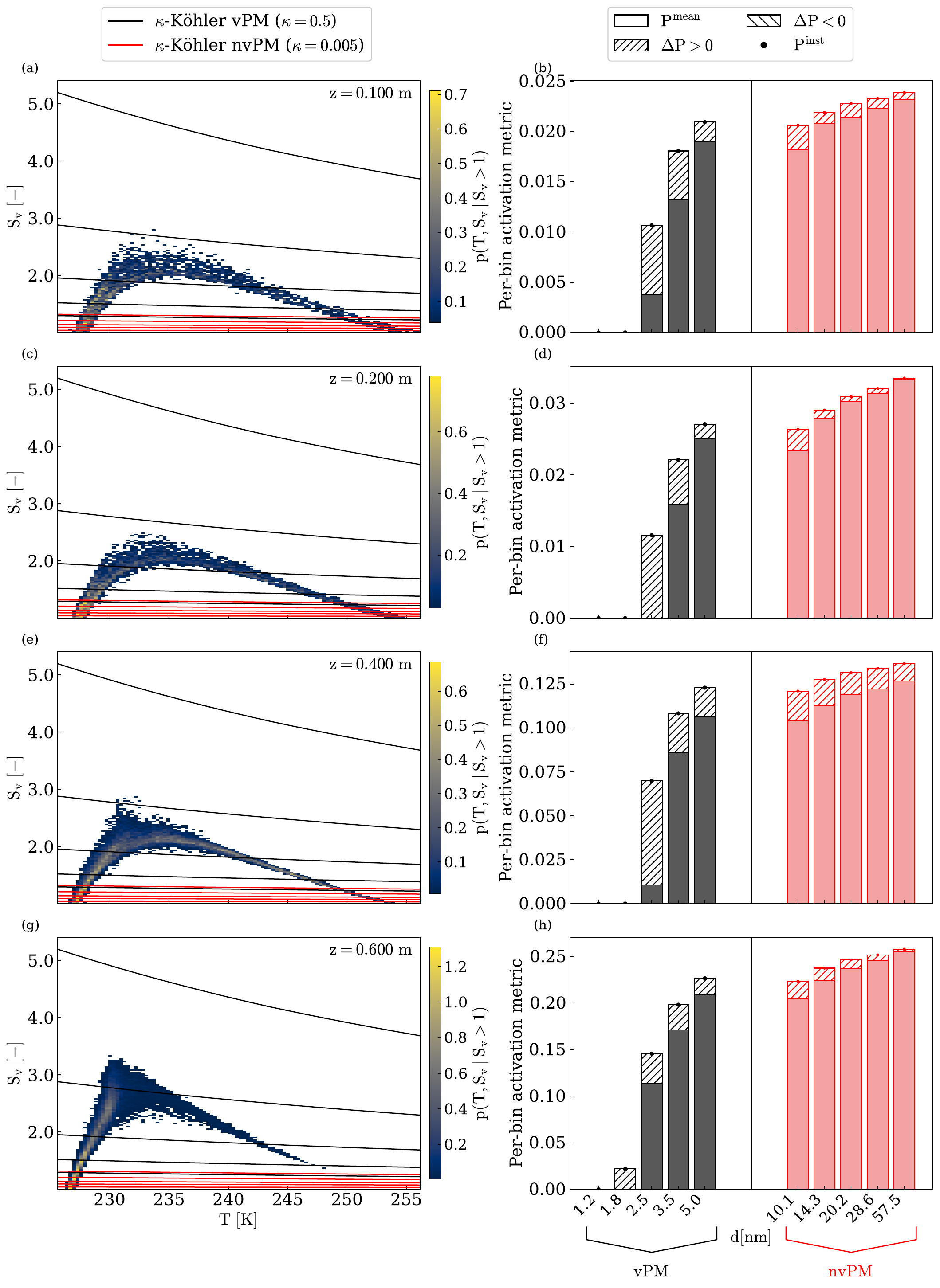}
  \caption{Joint-pdf (left column) and binwise contribution of time-averaged and turbulent fields to $\textnormal{vPM}$ and $\textnormal{nvPM}$ activation (right column) at four stramwise ($z$) locations for $\textnormal{H-S-3}$. The solid black and red lines in the left column are activation threshold values determined based on the $\kappa-$K\"{o}hler theory. In the right column the contribution of the turbulence field relative to the mean is quantified as detailed in section \ref{sec:methods_data_analysis}. Red colour represents  $\textnormal{nvPM}$ and black colour represents $\textnormal{vPM}$.}
  \label{fig:hs3_joint_pdf}
\end{figure}

To quantify the effect of turbulent fluctuations, the joint PDF  $p(T,S_v|S_v>1)$ is computed at four streamwise locations between the jet exit ($z=0 \textnormal{ m}$) and the region of maximum likelihood of INPs activation ($z=0.6\textnormal{ m}$). Contours of the joint PDF are shown in the left column of figure \ref{fig:hs3_joint_pdf}. This plot is the equivalent to \ref{fig:saturation_1d_pbe_hs3}, but now the joint PDF is shown at each $z$ location. As before, the curves of critical saturation against temperature for different diameters of dry particles are superimposed on the figure. 
For $\textnormal{nvPM}$ and for typical dry radii of soot particles \citep{Karcher2015}, the critical saturation (shown with red lines) has weak dependency on temperature and the activation threshold does not exceed $S_v\approx1.3$, which is just above the SAC ($S_v=1$). As can be seen, the majority of the events satisfy these conditions. On the other hand, for $\textnormal{vPM}$ larger values of liquid supersaturation are required for activation and $S_{v,c}$ increases with a decrease in dry particle radius (see black lines). The dependence on local temperature also increases with a decrease in $r_d$. The results indicate that the critical saturation ratios criterion for $\textnormal{vPM}$ activation are also met, although with a lower likelihood compared to $\textnormal{nvPM}$ activation thresholds. 

To quantify the role of turbulent fluctuations on the activation of the INPs, the metric $\Delta P$ introduced in section \ref{sec:methods_data_analysis} is used. In the right column of figure \ref{fig:hs3_joint_pdf}, the metric is plotted for $\textnormal{vPM}$ and $\textnormal{nvPM}$ and different dry core particle diameters. Values $\Delta P>0$ indicate activation enhanced by turbulence fluctuations, and vise-versa for $\Delta P<0$.  Relative to the mean, the contribution of turbulence fluctuations is negligible for the activation of the $\textnormal{nvPM}$ ($\Delta P\approx0$), while it is significant for the $\textnormal{vPM}$. In the region closer to the jet exit, $z<0.4\textnormal{ m}$, the activation of the smallest $\textnormal{vPM}$ ($d<3.5\textnormal{ m}$) is driven mainly by turbulence fluctuations, $\Delta P>0$, and it is more important than the contribution due to the time-average flow.

\begin{figure}
  \centering 
  \includegraphics[trim={0cm 0cm 0cm 0cm},clip,width=\textwidth]{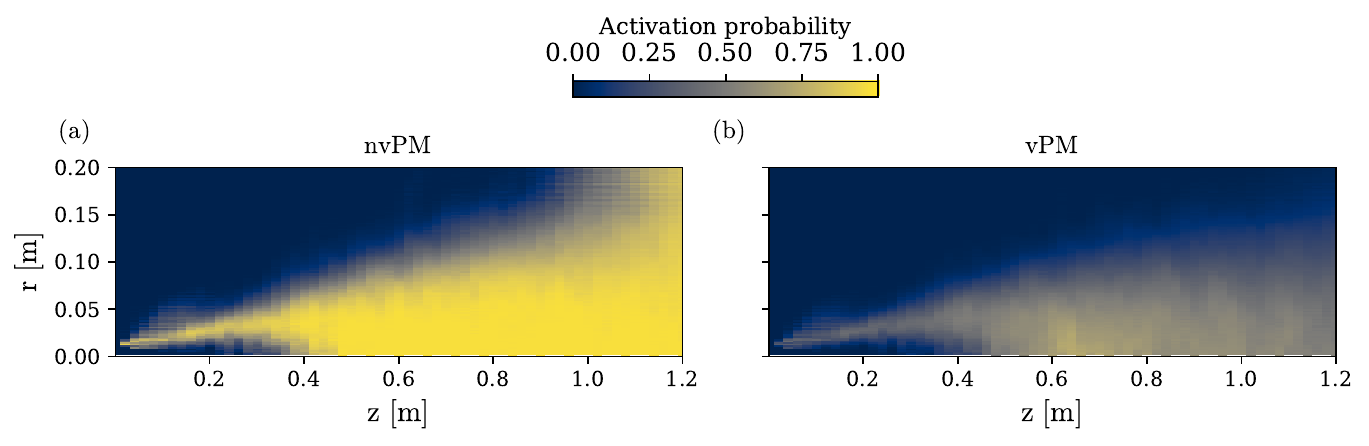}
  \caption{Meridional distribution of the activation probability for (\textit{a}) $\textnormal{nvPM}$ and (\textit{b}) $\textnormal{vPM}$ for $\textnormal{H-S-3}$.
  }
  \label{fig:activation_probability}
\end{figure}

To generate a spatial map for the activation of INPs, the diagnostic metrics introduced in section \ref{sec:methods_data_analysis} are used. Figure \ref{fig:activation_probability} shows the meridional  distribution of the activation probability $P_c^{\mathrm{inst}}(z)$ for $c=\textnormal{nvPM}$ (left) and $c=\textnormal{vPM}$ (right) for case $\textnormal{H-S-3}$ obtained from temporal and azimuthal ($\theta$) averaging. For both INPs classes, immediately downstream of the jet exit the activation probability is larger within the shear layer, where turbulence is more active (figure \ref{fig:hs3_statistics}). However, the peak value is closer to the jet centreline, where the water vapour concentration is greater, and further downstream ($z\approx0.65\textnormal{ m}$) where the jet has entrained colder ambient temperature and the time-averaged saturation ratio also peaks (figure \ref{fig:hs3_statistics}). Comparing panels (a) and (b) it becomes clear that  $\textnormal{nvPM}$ have higher probability to activate compared to the $\textnormal{vPM}$. This is due to the lower supersaturation threshold for $\textnormal{nvPM}$, as evidenced in the previous figure \ref {fig:hs3_joint_pdf}.  

\begin{figure}
  \centering 
  \includegraphics[trim={0cm 0cm 0cm 0cm},clip,width=\textwidth]{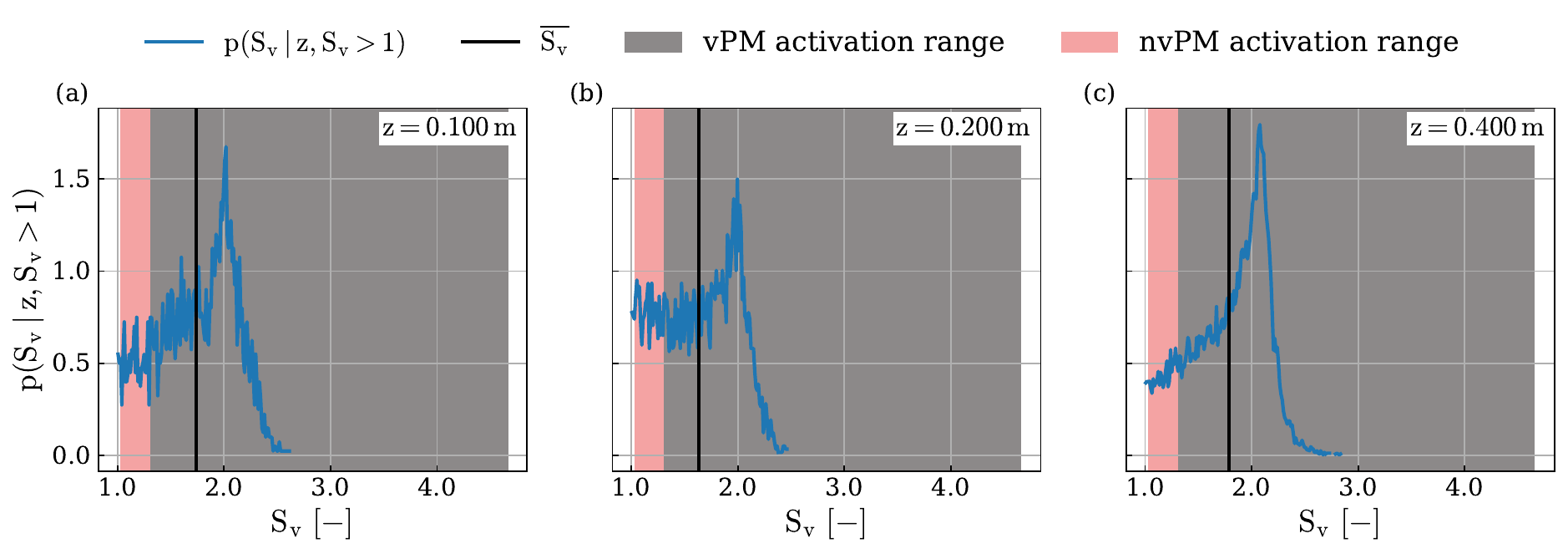}
  \caption{Conditional ($S_v>1$) pdf of liquid vapour saturation ratio at three streamwise ($z$) locations for $\textnormal{H-S-3}$. The vertical black solid line marks the time-average saturation ratio. The black and red shaded areas indicate the extent of the activation thresholds for $\textnormal{vPM}$ and $\textnormal{nvPM}$ respectively, based on the $\kappa-$K\"{o}hler theory.}
  \label{fig:hs3_pdf}
\end{figure}

The conditional pdf of the liquid vapour saturation ratio $p(S_v|S_v>1)$ is plotted in figure \ref{fig:hs3_pdf} at 3 streamwise locations. The extents of activation thresholds are also marked with different colours, red for  $\textnormal{nvPM}$ and gray for $\textnormal{vPM}$. The time-average saturation ratio is indicated with a vertical black line. The results further highlight the relative importance of mean flow and turbulence fluctuations on the activation of both INPs classes. In the area where the activation of the INPs mostly occurs within the jet shear-layer (figure \ref{fig:activation_probability}), the activation of the $\textnormal{nvPM}$ is driven by the mean flow, with the time-average saturation ratio ($\overline{S_v}$) to the right of activation threshold region. However, the peak of the pdf occurs at $S_v>\overline{S_v}$, and this confirms that the time-averaged flow is not sufficient to capture the activation of the smallest $\textnormal{vPM}$. Further downstream of the jet exit ($z>0.4\textnormal{ m}$, not shown), $\max(S_v)<\overline{S_v}$ , but the pdf is notably flat, with long, fat tails that extend over the region of $\textnormal{vPM}$ activation. Figure \ref{fig:hs3_pdf} further confirms the effect of turbulent fluctuations on ice crystal formation from $\textnormal{vPM}$. This is of notable importance for the modelling of contrails for SAF-fuelled aircraft.

\subsection{Effect of turbulent fluctuations on droplet and ice-crystal growth}
\label{sec:fluid_dynamics_crystal_growth}
\begin{figure}
  \centering 
  \subfloat[]{\includegraphics[trim={0cm 0cm 0cm 0cm},clip,width=0.5\textwidth]{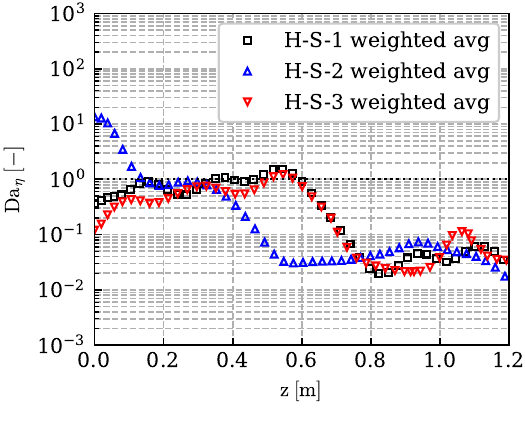}\label{fig:Da_eta_avg_hs}}    
  \hfill
  \subfloat[]{\includegraphics[trim={0cm 0cm 0cm 0cm},clip,width=0.5\textwidth]{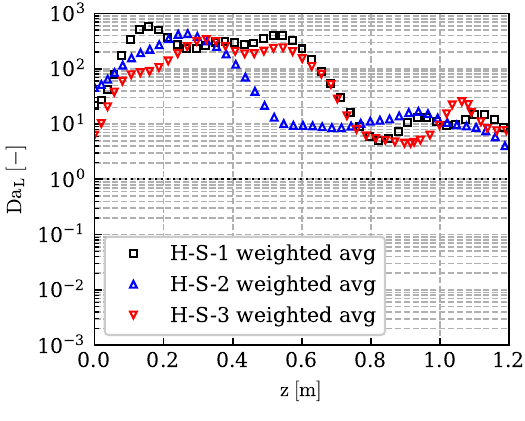}\label{fig:Da_L_avg_hs}}  
  \caption{Streamwise distribution of (mass flow averaged) Damk\"ohler number based on (\textit{a}) Kolmogorov timescale and (\textit{b}) eddy turnover time for the LES-PBE model of case $\textnormal{H-S}$. The black, dotted line marks $Da=1$.}
  \label{fig:hs_Da_les_pbe}
\end{figure}

Upon $\textnormal{vPM}$ activation, turbulence also affects the growth of water droplets and ice crystals. 
To quantify the interaction between turbulence and aerosol growth, it is customary \citep{Paoli2004,Tang2022} to introduce the non-dimensional Damk\"ohler number ($Da$), which is here defined as the ratio of the turbulent mixing time scale and particle growth time scale. For turbulent mixing, which occurs within the jet and  dominates the early stages of the jet dynamics, both the Kolmogorov time scale ($\tau_{\eta}$) and the eddy turnover ($\tau_{L}$) time are used,  
\begin{equation}
    \tau_{\eta}=\left(\frac{\nu}{\varepsilon}\right)^{1/2}, \; \tau_L = \frac{\frac{1}{2}\left( \overline{u^{\prime^2}} + \overline{v^{\prime^2}} + \overline{w^{\prime^2}} \right)}{\varepsilon}. 
    \label{eq:flow_timescales}
\end{equation}
Therefore two $Da$ can be defined,  
\begin{equation}
    Da_{\eta}=\frac{\tau_{\eta}}{\tau_g}, \; Da_L=\frac{\tau_L}{\tau_g} 
    \label{eq:Damkholer}
\end{equation}
The growth time scale ($\tau_g$) is determined from the ratio of the total particle volume to the total volumetric growth rate \citep{Baldyga1999}. Under the spherical particle assumption, 
\begin{equation}
    \tau_g = \frac{3(\frac{4}{3}\pi)^{2/3}\int_{v_0}^{v_{N_{bins}-1}}v \, n(v) \,dv}{4\pi \int_{v_0}^{v_{N_{bins}-1}} G(v) \, v^{-1/3} \, n(v) \,dv}.
    \label{eq:growth_timescale}
\end{equation}
Within the core of the jet plume the growth time scale ranges from approximately $10^{-3}\textnormal{ s}$ to $10^{-1}\textnormal{ s}$. In the region where a significant fraction of the $\textnormal{vPM}$ activates ($z\approx  0.5 \textnormal{ m}$ for $\textnormal{H-S-3}$, figure \ref{fig:activation_probability}b), figure \ref{fig:Da_eta_avg_hs} shows that $Da_{\eta}\approx1$, which means that the growth timescales are comparable to the Kolmogorov timescales. As the jet approaches ambient conditions driven by both dilution and aerosol particle feedback, the kinetics slows down, but the growth time scale remains comparable with the eddy turnover time scale ($Da_L\approx10$, figure \ref{fig:Da_L_avg_hs}). The activation of a larger number of $\textnormal{vPM}$ for $\textnormal{H-S-2}$ compared to $\textnormal{H-S-3}$, leads to a faster consumption of water vapour and smaller growth timescale leading to streamwise distributions of $Da_{\eta}$ and $Da_L$ that deviate from the other two cases  $\textnormal{H-S-1}$ and $\textnormal{H-S-3}$. Overall, the analysis provides further confirmation that in the region of $\textnormal{vPM}$ activation strong dynamic coupling between the aerosol kinetics and turbulence occurs. 

\vspace*{-5pt}
\section{Conclusions}\label{sec:conclusions}
This work considers the effect of turbulent jet dynamics on the activation of ice nucleating particles (INPs) during the early stages of contrail formation. A detailed ice-microphysics model is incorporated into a compressible LES solver, resulting in a coupled LES-PBE framework that accounts for the activation of both non-volatile (soot-type) and volatile (sulphur-type) particles. The latter play a key role in contrail formation from aircrafts powered by SAF. The computational model is applied to an experimental configuration that reproduces the thermodynamic conditions encountered at cruise altitude. 

The assessment of the numerical simulations with experimental data shows that the LES can capture well the jet-core cooling rate, with a root-mean-squared error of approximately $2.5 \textnormal{ K}$ in the radial temperature profiles within the core (further away comparison is not possible due to lack of information on wall temperature boundary conditions). The accuracy of the LES-PBE model is assessed for both conventional (soot-rich) fuel and SAF. Although there is no experimental information on $\textnormal{vPM}$ for the SAF cases, it is shown that the LES-PBE simulations can bracket the experimental data of ice properties when $\textnormal{vPM}$ is also introduced using judicious values for number density and size distribution. 

For soot-rich fuels, the 1D model proves sufficiently adequate for the prediction of particle activation and ice crystal formation and growth. For the SAF case however, the 1D-PBE model significantly underestimated ice crystal concentrations at the end of the jet-regime, due to the lack of $\textnormal{vPM}$ activation. Statistical analysis of the thermodynamic and aerosol quantities from the LES-PBE results indicated that the $\textnormal{vPM}$ activation is largely driven by supersaturation fluctuations that naturally occur within the jet plume due to turbulence. This may be an important limitation of current simplified contrail models, which is likely to propagate into the uncertainty of contrail-induced radiative forcing computed with global models.  

Overall, building upon the two-way coupled LES-PBE model validation with experimental data, this work provides the motivation to understand at a more fundamental level the activation pathways of $\textnormal{nvPM}$ and $\textnormal{vPM}$ and the subsequent growth of ice-crystals under realistic flight scenarios and aero-engine configurations.

\section*{Acknowledgements}
The authors acknowledge the UK Turbulence Consortium (funded by EPSRC grant EP/R029326/1) for providing access to the UK supercomputing facility ARCHER2. Computational resources were also provided by the Imperial College Research Computing Service (\url{http://doi.org/10.14469/hpc/2232}). This work also used the Cirrus UK National Tier-2 HPC Service at EPCC (\url{http://www.cirrus.ac.uk}) funded by the University of Edinburgh and EPSRC (EP/P020267/1). L.B.\ acknowledges support from the EU Horizon Europe MSCA (CASSANDRA, GA~101263613). The authors thank Dr. A. Fagan and NASA for permission to use their experimental data. L.B. acknowledges Dr. D. Fredrich and G. Efstathiou for helpful discussions on the development of the numerical model. All authors thank Prof. W. P. Jones for making the BOFFIN code available.

\section*{Funding}
This research received financial support of UKRI-NERC (grant reference number: NE/Z503861/1)

\section*{Declaration of interests}
The authors report no conflict of interest.

\section*{Data Availability Statement}
Raw data are available from the corresponding author upon request.

\section*{Author ORCIDs}
L. Boscagli, \url{https://orcid.org/0000-0002-8121-4208};
 S. Rigopoulos \url{https://orcid.org/0000-0002-0311-2070}; G. Papadakis \url{https://orcid.org/0000-0003-0594-3107}

\appendix

\section{Equations of ice microphysics model}\label{app:ice_microphysics}
This section summarises the thermodynamic expressions for the liquid and ice saturation curves and the latent heat release due to phase change. A concise description of the switching criterion between condensational and depositional growth of particles based on homogeneous freezing temperature is also provided.

\subsection{Thermodynamic correlations }\label{app:thermo_correlations} 
The liquid ($p_{v}^{liq}$) and ice ($p_{v}^{ice}$) saturation partial pressures are given by \citep{Murphy2005},
\begin{equation}
\begin{aligned}
p_{\text{v}}^{\text{liq}}  = \exp\!\Bigg(&
a_{l,1}
- \frac{a_{l,2}}{T}
- a_{l,3} \ln(T)
+ a_{l,4}\, T 
+ \tanh\!\left(a_{l,5} (T - 218.8)\right)\\
&\Big(
a_{l,6}
- \frac{a_{l,7}}{T}
- a_{l,8} \ln(T)
+ a_{l,9}\, T
\Big)
\Bigg),
\end{aligned}
\label{eq:pv_liq}
\end{equation}

\begin{equation}
\begin{aligned}
p_{\text{v}}^{\text{ice}} = \exp\!\Big(
& a_{i,1}
- \frac{a_{i,2}}{T}
+ a_{i,3} \ln(T)
- a_{i,4}\, T
\Big),
\end{aligned}
\label{eq:pv_ice}
\end{equation}
where $a_{l,1}=54.842763$, $a_{l,2}=6763.22$, $a_{l,3}=4.210$, $a_{l,4}=0.000367$, $a_{l,5}=0.0415$, $a_{l,6}=53.878$, $a_{l,7}=1331.22$, $a_{l,8}=9.44523$, $a_{l,9}=0.014025$, and $a_{i,1}=9.550426$, $a_{i,2}=5723.265$, $a_{i,3}=3.53068$, $a_{i,4}=0.00728332$. The expressions for $p_{\text{v}}^{\text{liq}}$ and $p_{\text{v}}^{\text{ice}}$ are valid for $123\textnormal{K}<T<332\textnormal{K}$ and $110\textnormal{K}<T<273.16\textnormal{K}$, respectively.

The latent heat of condensation and deposition is computed from \citep{Yau1996}, 
\begin{equation}
l_{c/d}(T) =
\begin{cases}
\begin{aligned}
\displaystyle
2.5 \times 10^{6}
- 2.36 \times 10^{3}\,(T - 273.15)
+ 1.6\,(T - 273.15)^2 \\
\displaystyle
- 0.06\,(T - 273.15)^3,
\end{aligned}
& T > T_{\mathrm{frz}}, \\[12pt]

\begin{aligned}
\displaystyle
2.8346 \times 10^{6}
- 340.0\,(T - 273.15) \\
\displaystyle
- 10.46\,(T - 273.15)^2,
\end{aligned}
& T \le T_{\mathrm{frz}},
\end{cases}
\end{equation}
where $T_{\mathrm{frz}}$ is the freezing temperature (see below). 

\subsection{Freezing temperature criterion}\label{app:freezing_criterion}
The homogeneous freezing temperature criterion \citep{Karcher2015} for supercooled water droplets is used to switch between condensational and depositional growth. The criterion requires the computation of the  freezing temperature,  
\begin{equation}
    T_{frz} = \begin{cases}0 \textnormal{ K}, & \text{if } LWV=0\textnormal{ m}^{3}\\ \frac{1}{a_1}\left[\ln\left(\frac{10^{-6}a_1 \dot{T}}{LWV}\right)-a_2\right],&\text{otherwise.}\end{cases}
\end{equation}
where $a_1=-3.5714 \textnormal{ K}^{-1}$, $a_2=858.719$, and $\dot{T}$ is the jet cooling rate, which is determined based on \cite{Karcher2015}. $LWV$ is the available liquid water volume in the droplet, and it is determined as the difference of the spherical volume of the (active) supercooled water droplet and the volume of the dry (inactive) core. In order to model the transport of multiple particles with a single PBE, the dry core radius is only accounted for the smallest active particle, and ignored for the larger particles. This is a reasonable approximation because upon activation and initial condensational growth, the size of the supercooled water droplets is significantly larger than the size of the dry core, and therefore it is reasonable to use only the wet particle radius to compute $LWV$ \citep{Karcher2015}. Thus, the homogeneous freezing criterion is summarised as,
\begin{equation}
    \frac{dm}{dt} = 
    \begin{cases}
    \text{Eq. } \eqref{eq:water_dmdt} & \text{if } T>T_{frz}\\ 
    \text{Eq. } \eqref{eq:ice_dmdt} & \text{if } T\leq T_{frz}
    \end{cases},
\end{equation}
where $T$ is the local gas (and particle) temperature.

\section{Assessment of PBE grid resolution}
\label{app:pbe_grid_refinement}
\begin{figure}
  \centering 
  \subfloat[]{\includegraphics[trim={0cm 0cm 0cm 0cm},clip,width=0.5\textwidth]{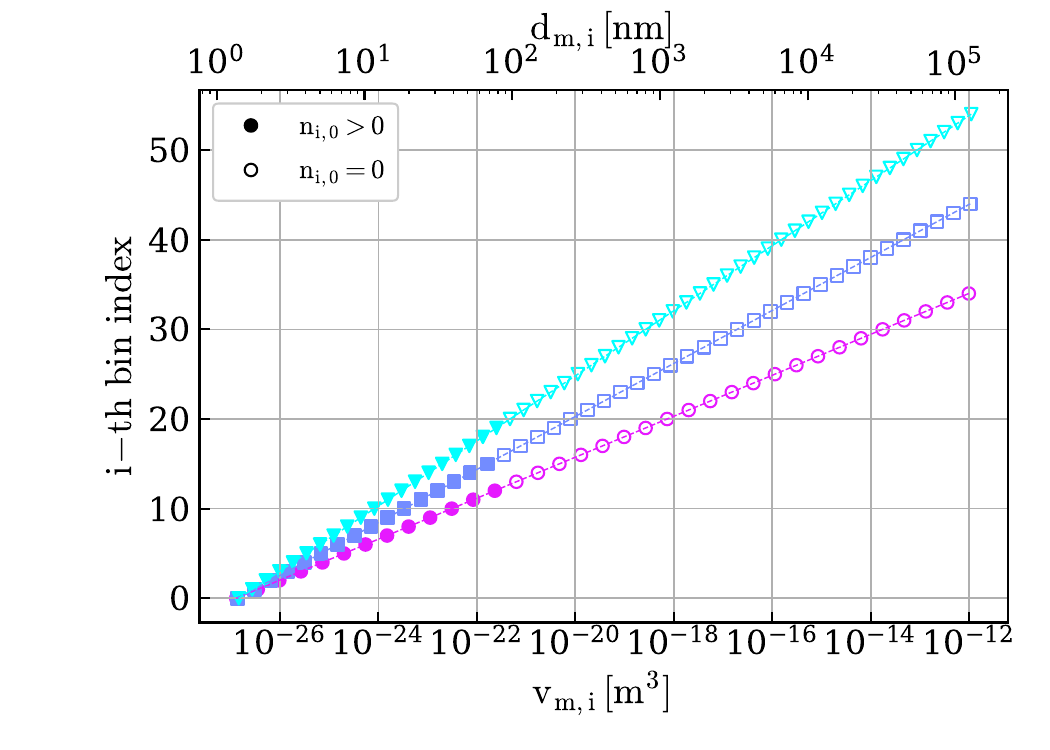}\label{fig:pbe_grid_res}}
  \hfill 
  \subfloat[]{\includegraphics[trim={0cm 0cm 0cm 0cm},clip,width=0.5\textwidth]{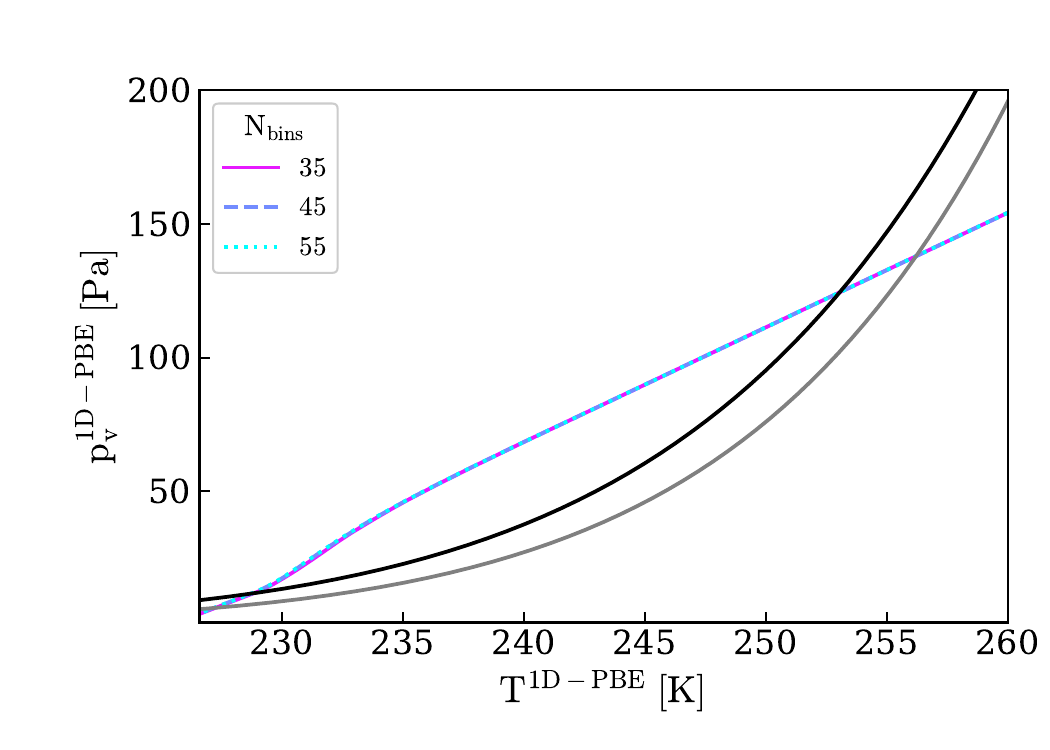}\label{fig:pv_grid_ref}}
  \hfill 
  \subfloat[]{\includegraphics[trim={0cm 0cm 0cm 0cm},clip,width=0.5\textwidth]{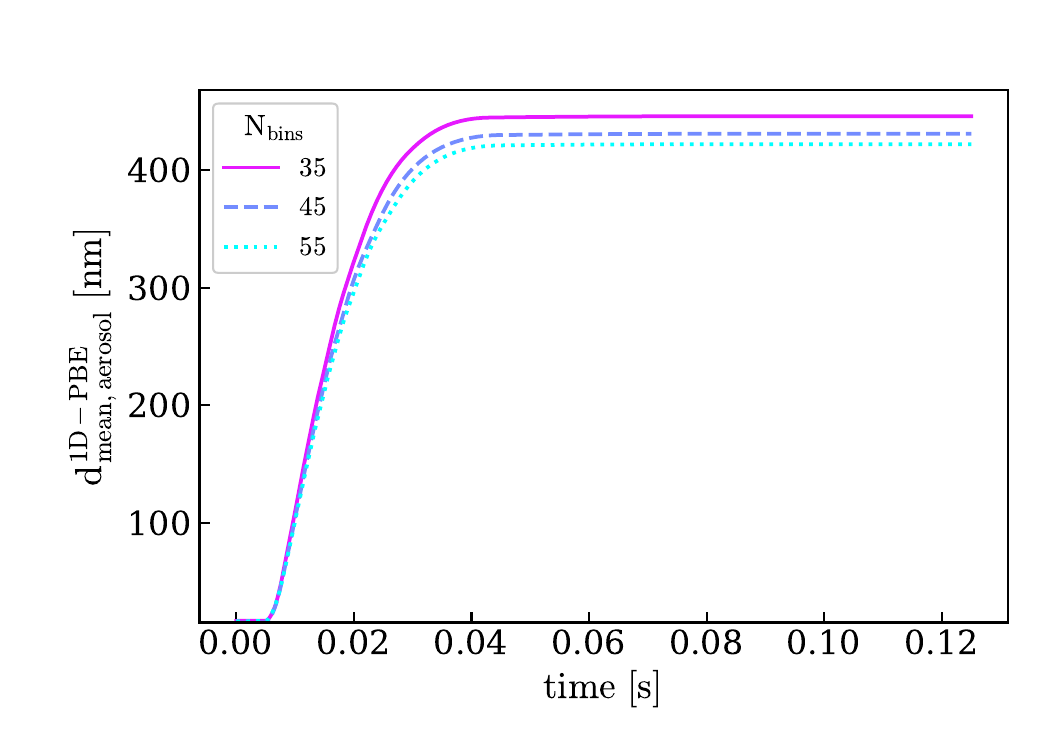}\label{fig:D_grid_ref}} 
  \hfill 
  \subfloat[]{\includegraphics[trim={0cm 0cm 0cm 0cm},clip,width=0.5\textwidth]{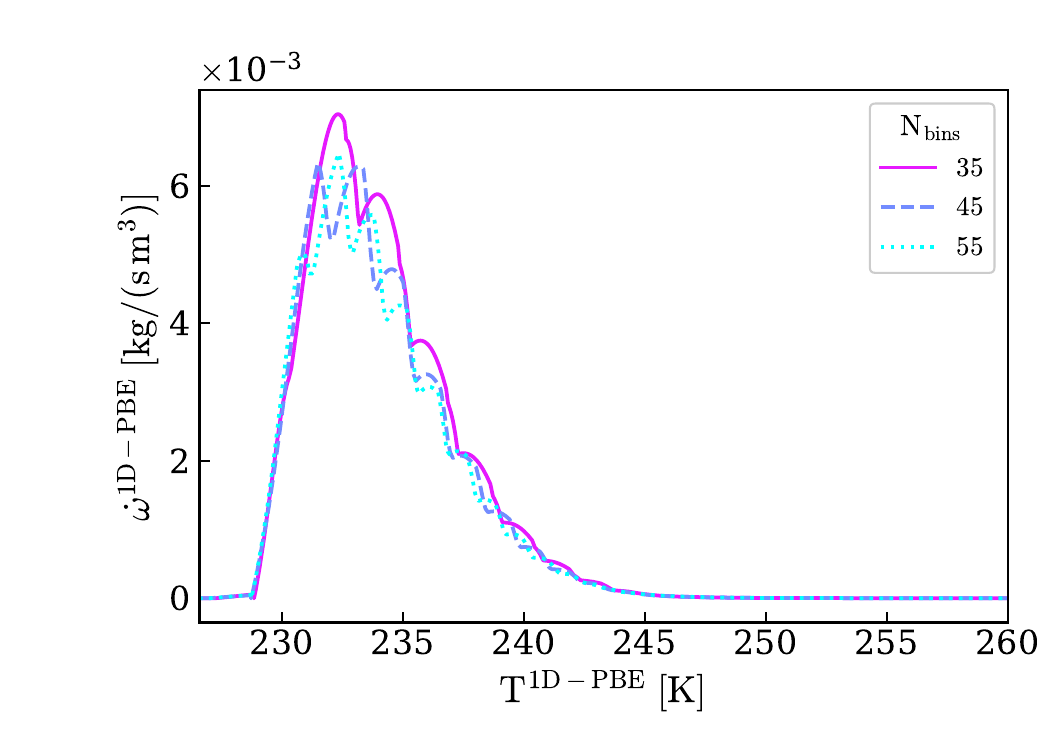}\label{fig:msource_grid_ref}}  
  \caption{1D-PBE model investigation of the effect of PBE grid refinement on (\textit{a}) bins distribution, (\textit{b}) water vapour partial pressure, (\textit{c}) aerosol mean diameter and (\textit{d}) water vapour consumption rate. Filled and open symbols in (\textit{a}) indicate the initial bins with number concentration greater than and equal to zero, respectively. The grey and black solid lines in (\textit{b}) indicate the ice and liquid vapour saturation curves, respectively.}
  \label{fig:pbe_grid_refinement}
\end{figure}

The resolution of the PBE grid is assessed for case $\textnormal{H-S-3}$ with initial ambient and jet temperature conditions and INPs characteristics detailed in tables  \ref{tab:les_pbe_combined} and \ref{tab:inps_details} respectively. The slope of the mixing line is $G^{\mathrm{mix}}=4.28$, determined from the centreline LES-PBE data for $T_a=225 \textnormal{ K}$ as detailed in section \ref{sec:1d_pbe_method}. Three bin numbers ($N_{bins}=35,45,55$) were used to discretise the  PBE equation \eqref{eq:pbe_step2}. The bin distributions are shown in figure \ref{fig:pbe_grid_res}. 

Overall, it is found that water vapour partial pressure (figure \ref{fig:pv_grid_ref}) and mean aerosol diameter ($d_{mean,aerosol}$, figure \ref{fig:D_grid_ref}) are only slightly affected by the PBE grid resolution, with the relative, percentage change in maximum water vapour consumption rate among $N_{bins}=35$ and $N_{bins}=55$ that remains below 10\% (figure \ref{fig:msource_grid_ref}). For the purposes of this study, which focuses on the fluid dynamic mechanisms driving \textnormal{vPM} activation, $N_{bins}=35$ is selected as it provides a good balance between computational cost and accuracy.

\bibliographystyle{jfm}
\bibliography{jetzero}

\end{document}